\documentclass[twoside,a4paper,11pt]{amsart}

\usepackage[english]{babel}
\usepackage{amsmath}
\usepackage{amssymb}
\usepackage{url}
\usepackage{amsthm}
\usepackage{amsxtra}
\usepackage{mathrsfs}
\usepackage{fancyhdr}
\usepackage{pdfsync}
\usepackage{hyperref}

\newtheorem{theorem}{Theorem}
\newtheorem{lemma}{Lemma}[section]
\newtheorem{proposition}[lemma]{Proposition}

\newtheorem{definition}[lemma]{Definition}

\theoremstyle{definition}
\newtheorem{remark}[lemma]{Remark}

\numberwithin{equation}{section}

\newcommand{\pd}[2]{\frac{\partial {#1}}{\partial {#2}}}
\newcommand{\beq}{\begin{equation}}
\newcommand{\eeq}{\end{equation}}
\newcommand{\be}{\begin{equation*}}
\newcommand{\ee}{\end{equation*}}

\newcommand{\RE}{\mathbb R}

\newcommand{\CO}{\mathbb C}
\newcommand{\NA}{\mathbb N}
\newcommand{\II}{\mathbb I}

\newcommand{\Hbb}{\mathbb{H}}
\newcommand{\BB}{\mathscr B}

\newcommand{\OO}{\mathcal{O}}

\newcommand{\HH}{\mathbb{H}}
\newcommand{\GG}{\mathcal{G}}

\newcommand{\Ran}{\operatorname{Ran}\,}

\newcommand{\ve}{\varepsilon}
\newcommand{\al}{\alpha}

\newcommand{\ga}{\gamma}
\newcommand{\La}{\Lambda}
\newcommand{\la}{\lambda}
\newcommand{\de}{\delta}

\renewcommand{\Im}{\operatorname{Im}\,}

\newcommand{\diff}{\operatorname{d}\!}

\newcommand{\WW}{\mathcal{W}}

\newcommand{\hgraf}{\Hbb_{\mathcal{G}}}
\newcommand{\NN}{\mathbb{N}}
\newcommand{\PP}{\mathbb{P}}
\newcommand{\hmet}{\mathfrak{h}}
\newcommand{\gmet}{\mathfrak{g}}
\newcommand{\deve}{{\de,\ve}}
\newcommand{\defve}{{\de/\ve}}
\newcommand{\hatpsiuu}{{\hat\psi_{1,\ve}}}
\newcommand{\hatpsidd}{{\hat\psi_{2,\ve}}}
\newcommand{\hatpsijj}{{\hat\psi_{j,\ve}}}
\newcommand{\hatpsiv}{{\hat\psi_{v,\ve}}}

\newcommand{\puu}{{p_1}}
\newcommand{\pdd}{{p_2}}
\newcommand{\quu}{{q_{1,\ve}}}
\newcommand{\qdd}{{q_{2,\ve}}}
\newcommand{\xiuu}{{\xi_{1,\ve}}}
\newcommand{\xidd}{{\xi_{2,\ve}}}
\newcommand{\hatPsi}{{\hat\Psi_\ve}}
\newcommand{\psistar}{{\hat \psi^*_\ve}}

\newcommand{\hdec}{{{\bf h}^{dec}}}
\newcommand{\rdec}{{{\bf r}^{dec}}}
\newcommand{\halpha}{{{\bf h}^{\al_1\al_2}}}
\newcommand{\ralpha}{{{\bf r}^{\al_1\al_2}}}

\newcommand{\chide}{{\chi_{\de,n}}}

\newcommand{\Hunoring}{\mathring{H}^1_\deve}
\newcommand{\Hduering}{\mathring{H}^2_\deve}

\newcommand{\vv}{\underline{v}}

\title[]{Graph-like asymptotics for the Dirichlet Laplacian in connected tubular domains}

\author[]{Claudio Cacciapuoti}

\begin{document}

\maketitle
\begin{center}
\address{DiSAT, Sezione di Matematica, Universit\`a dell'Insubria\\  Via Valleggio 11, 22100
Como, Italy}\\

\email{claudio.cacciapuoti@uninsubria.it}
\end{center}

\begin{abstract}
We consider the Dirichlet Laplacian in a waveguide of uniform width and infinite length  which is ideally divided into three parts: a ``vertex region'', compactly supported and with non zero curvature, and two ``edge regions'' which are semi-infinite straight strips. We make the waveguide collapse onto a graph by  squeezing the edge regions to half-lines and the vertex region to a point. In a setting in which the ratio between the  width of the waveguide and the longitudinal extension of the vertex region goes to zero,  we prove the convergence of the operator to a selfadjoint realization of the Laplacian on a two edged graph. In the limit operator, the boundary conditions in the vertex depend on the spectral properties of an effective one dimensional Hamiltonian associated to the vertex region.  
\end{abstract}

\begin{footnotesize}
\end{footnotesize}

\section{Introduction}

Metric graphs are objects of great interest as simple but highly not
trivial tools to investigate questions covering several topical issues
in pure and applied mathematics. They  offer a twofold advantage: they
are quite simple objects (in many cases formulas for the relevant
quantities can be explicitly computed), showing at  the same time non
trivial features. We evade  the duty of giving  an exhaustive list of
applications of metric graphs and  limit ourselves to point out several recent volumes on the subject  \cite{berkolaiko-carlson-fulling-kuchment:06, berkolaiko-kuchment:13, exner-keating-kuchment-sunada-teplyaev:08, mugnolo:14}, where an extensive list of references can be found.

Our work is addressed to investigate a long standing problem within the
theory of metric graphs, namely, to find a rigorous justification for
the use of graph models to approximate dynamics in  networks of thin
waveguides. In many applications the problem is  reduced to compare, in
some suitable sense,  the Laplacian in a squeezing  network of thin tubes  with a
selfadjoint realization of the Laplacian on the graph. A difficult
feature of the problem originates from the fact that requiring
selfadjointness does not fix univocally the operator
on  the graph. \\ Each  selfadjoint realization of the Laplacian on a graph
is associated to certain boundary (or matching) conditions in the
vertices. There are different ways  to express the most general
selfadjoint boundary conditions, see, e.g., \cite{cheon-exner-turek:ap10, kostrykin-schrader:99, kuchment:04}, for the
sake of simplicity let us specify one of them for the case of a
star-graph $\GG$, that is a graph with $N$ edges of infinite length and
one   vertex $v$. We denote by $\hgraf$ the Hilbert space naturally
associated to $\GG$, a function $x\in\hgraf$ is a $N$-component vector valued function, $x=(x_1,...,x_N)$, such that $x_j\in L^2((0,\infty))$ is the component of the wave function associated to the $j$-th edge of the graph. In each edge the vertex $v$ is identified with the origin of the half-line $[0,\infty)$.  Consider an orthogonal projector $\Pi$  in $\CO^N$, and a
selfadjoint operator $\Theta$ in $\Ran(\Pi^\perp)\subseteq \CO^N$, with
$\Pi^\perp=\II-\Pi$. We denote by $-\Delta_\GG^{\Pi,\Theta}$ the
selfadjoint realization of the Laplacian  in $\hgraf$
associated to the couple of operators  $(\Pi,\Theta)$. Let
$x=(x_1,...,x_N)$ be a  vector in the domain of
$-\Delta^{\Pi,\Theta}_\GG$, then  $x_j\in H^2((0,\infty))$ and  $x$ must
satisfy the boundary conditions  
\be
\Pi x(v)=0\;,\quad \Pi^\perp x'(v) +\Theta\Pi^\perp x(v)=0\,,
\ee
where we denoted by $x(v)$ and $x'(v)$ the vectors in $\CO^N$ defined by $x(v)\equiv\big(x_1(0),...,x_N(0)\big)$ and $x'(v)\equiv\big(x_1'(0),...,x'_N(0)\big)$,  moreover  $-\Delta_\GG^{\Pi,\Theta} x= (-x_1'',...,-x_N'')$. 

Among the possible realizations of $-\Delta_\GG^{\Pi,\Theta}$ we mention: 
\begin{itemize}
\item[-] The Dirichlet (or decoupling) Laplacian: it is defined by the boundary condition $x(v)=0$, which corresponds to 
the choice $\Pi=\II$, in this case $\dim[\Ran(\Pi^\perp)]=0$.
\item[-] The weighted Kirchhoff Laplacian: it is defined by the conditions $\al_j x_i(0)=\al_i x_j(0)$, $i,j=1,...,N$, and $\sum_{i=1}^N\bar \al_ix_i'(0)=0$, with $\al_i\in\CO$, $\sum_{i=1}^N|\al_i|^2\neq0$. This conditions correspond to the choice 
\beq
\label{PiTh}
(\Pi)_{ij}=\de_{ij}-\frac{\al_i\bar \al_j}{\sum_{k=1}^N|\al_k|^2}\,,\quad i,j=1,...,N\,,\qquad \Theta=0\,,
\eeq
where $\de_{ij}$ denotes the Kronecker symbol. In this case
	 $\dim[\Ran(\Pi^\perp)]=1$.  When all the constants $\al_j$ are
	 equal the operator defined by the choice \eqref{PiTh} is
	 called  Kirchhoff (or standard) Laplacian and the boundary
	 conditions read $x_1(0)=...=x_N(0)$ and
	 $\sum_{i=1}^Nx_i'(0)=0$. 
\end{itemize}
A central problem in using  operators in the family
$-\Delta_\GG^{\Pi,\Theta}$ to approximate the dynamics generated by the
Laplacian in a network of thin tubes squeezing to a graph is to
understand which boundary conditions in the vertex arise in limit (see,
e.g., \cite{exner-kovarik:15, exner:11, post:12} for a review on this topic). It
turns out that this problem strongly depends on what kind of Laplacian
is taken in the squeezing network. 

With Neumann conditions on the boundary of the squeezing network  (or if
one considers a network without boundary see, e.g., \cite{exner-post:05,
post:06}) the problem is in some sense easier. Under general assumptions
on the network properties (see the discussion in \cite{exner-post:05}),
the limit operator on the graph is characterized by weighted Kirchhoff conditions
in the vertex, with $\al_i$ real and  the ratio $\al_i/\al_j$ being related to the relative radius of network's edges. \\ 
In the 
setting of a Neumann network a first result  can be traced back to the work \cite{deverdiere:86}, where the convergence of the eigenvalues is used to show that in dimension three or higher it is possible to construct a compact Riemannian manifold such that the  first nonvanishing eigenvalue of the Laplacian  has arbitrary multiplicity. Later on, the problem has been studied from the point of view of the convergence of
stochastics processes, see 
\cite{freidlin-wentzel:93} (see also \cite{albeverio-kusuoka:pp10}, for 
recent results in this direction in a setting with Dirichlet conditions). In \cite{exner-post:05, kuchment-zeng:01,  rubinstein-schatzman:01}  the convergence of eigenvalues  was analyzed  while in \cite{post:06} the convergence of operators  in norm resolvent sense was proved (see also \cite{saito:00, saito:01}). We also mention the works  \cite{bonciocat:08, kuroda:11} in which the problem is analyzed by using the theory of Gromov-Hausdorff convergence of manifolds and the techniques developed in \cite{kuwae-shioya:03}. \\
In the setting with Neumann boundary conditions we quote  also: the work \cite{exner-post:07-1} in which the convergence of resonances is analyzed and  the work \cite{exner-post:09} where the presence of a scaled potential supported in the vertex region is taken into account. It is also worth pointing out that there exist works addressed to the analysis of the corresponding nonlinear problem: see, e.g.,  \cite{raugel:95}, and references therein; \cite{kosugi:00} in which the convergence of semilinear elliptic equations is studied;  and \cite{ueker-ea:15} for the study of the nonlinear Schr\"odinger equation.

The case of a Dirichlet network squeezing to a graph is more complicated
and not yet completely understood. One main difficulty is related to the
fact that as the width of the network squeezes to zero the energy
functional diverges as the inverse of the square of the width. Hence, to
get a meaningful, limit a renormalization of the energy is
needed. Moreover geometrical perturbations  can modify substantially the
spectral properties of the Laplacian in the network, a well known phenomenon is, for example,  the appearing  of isolated eigenvalues in curved waveguides, see  \cite{duclos-exner:95, exner-kovarik:15, ESjmp}. Similar problems
arise when the confinement is obtained through an holonomic constrain,
see, e.g., \cite{dellantonio-tenuta:jmp06}.  \\ 
In \cite{post:05} it is 
shown that in a Dirichlet network, if the volume of the vertex
region  is small enough, then the  boundary conditions in the vertex are of
Dirichlet type. Nevertheless it is known that more general assumptions
cannot exclude different limits. In particular it is understood that
weighted Kirchhoff conditions of the form described above may arise
whenever an effective Hamiltonian associated to the vertex region
exhibits eigenvalues at the threshold of the transverse energy see,
e.g., \cite{grieser:07, harmer-pavlov-yafyasov:07,
molchanov-vainberg:06-2},  and 
\cite{dellantonio-costa:10,dellantonio-michelangeli:15}, in which these ideas are further exploited
and the occurrence of non homogenous terms in the weighted Kirchhoff
condition is pointed out. 

In a previous work \cite{albeverio-cacciapuoti-finco:07} we considered the Laplacian with Dirichlet conditions on the boundary of  a planar waveguide of uniform width collapsing onto a two edged graph. We  proved that two possible limits can  arise: 
\begin{itemize}
\item[-] In one case, which we call generic, the limit operator on the graph is defined by the Dirichlet condition in the vertex.
\item[-] In a second case, which we call non-generic, the limit operator on the graph is defined by a weighted Kirchhoff condition in the vertex. The occurrence of the non-generic case is related to the existence of a zero energy resonance for an effective one dimensional Hamiltonian associated with the squeezing waveguide.
\end{itemize}
A crucial hypothesis within the model analyzed in
\cite{albeverio-cacciapuoti-finco:07} is the assumption that there exist two
scales of energy: a short one associated to the width of the waveguide,
$\de$; and a large one associated to the longitudinal extension of the
vertex region, $\ve$, with $\de<\ve^{5/2}$. This assumption implies that
the dynamics in the waveguide is adiabatically separated into a
``fast''  component associated to the transverse coordinate and a
``slow'' component associated to the longitudinal coordinate, making it
possible to analyze  the  problem in two steps. First the dynamics is
reduced to  a one dimensional one by  projecting  onto the transverse eigenfunctions and by taking the limit $\de\to0$ (in norm resolvent sense); then the operator on the graph is  obtained  by taking the limit $\ve\to0$ (in norm resolvent sense) of the effective one dimensional Hamiltonian. For the analysis of a similar problem in three dimensional waveguides with methods of Gamma-convergence we mention the paper  \cite{oliveira:rxv10}, see also \cite{bouchitte-mascarenhas-trabucho:esaim07}.


Aim of the present work is to give a better understanding of the result obtained in \cite{albeverio-cacciapuoti-finco:07} in view of applications to multi-edged graphs. We review the problem separating explicitly the edges and vertex regions in the waveguide. According to the ideas in \cite{dellantonio-costa:10,dellantonio-michelangeli:15, grieser:07, harmer-pavlov-yafyasov:07, molchanov-vainberg:06-2}, we reinterpret the existence of the zero energy resonance (which leads to coupling conditions in the vertex) as an eigenvalue at the energy of the transverse modes for an effective Hamiltonian associated to the vertex region. Moreover in the generic case we obtain better estimates (as compared to \cite{albeverio-cacciapuoti-finco:07}) for the rate of convergence ($\de<\ve^{3/2}$, instead of $\de<\ve^{5/2}$).

Related to the topic of our paper are also several works which concern
the approximation 
of generic matching conditions in the vertex by  scaled  Schr\"odinger
operators on the graph, see for example \cite{cheon-exner-turek:ap10, manko:12} and the  review
\cite{exner:11}. We also point out  the  paper
\cite{golovaty-hryniv:09} (and  
references therein) 
concerning the analysis of the convergence of Schr\"odinger operators in
dimension one to $\delta^\prime$-type point interactions.

Here is a short summary of our paper. In section \ref{model} we define our model for
the waveguide as the union of three parts: the two edges and the vertex
region. Moreover we define the Dirichlet Laplacian in the waveguide. In
section \ref{sec2} we present our main results (Theorem \ref{resconv}
and Theorem \ref{asymbeh}). Sections \ref{secth1} and \ref{secth2} are
devoted to the proofs of the main theorems. In section \ref{sec:graph}
we explain how to interpret the result stated in Theorems \ref{resconv}
and \ref{asymbeh} in terms of the convergence (in norm resolvent sense)
to an operator on the graph, see Theorem \ref{th3}. 
We close the paper
with section \ref{conc} in which we
summarize our results and
 point out some generalizations.

\section{The model}
\label{model}

We consider a smooth  waveguide of uniform width $\de>0$ embedded in $\RE^2$. We assume that the waveguide can be decomposed into  three parts: two straight edges $E_{1,\de}$ and $E_{2,\de}$  and one  vertex region  $V_{\de,\ve}$, $\ve >0$ being a parameter characterizing the ``longitudinal extension'' of the vertex region. Throughout the paper we shall always assume that $\de/\ve\leqslant 1$. Each edge $E_{j,\de}$, $j=1,2$, can be identified with the manifold $(E,\hmet_{\de})$ where $E:=(0,\infty)\times(0,1)$ and $\hmet_\de$ is the metric
\be
\hmet_\de:=\diff s^2+\de^2 \diff u^2\,,
\ee
where $s\in(0,\infty)$ and $u\in (0,1)$. The vertex region can be identified with the manifold $(V,\gmet _{\de,\ve})$ where $V:=(-1,1)\times(0,1)$ and $\gmet_{\de,\ve}$ is the metric
\be
\gmet_{\de,\ve}:=\ve^2g_{\de/\ve}\diff s^2+\de^2\diff u^2
\ee
with $s\in(-1,1)$ and $u\in(0,1)$. The function $g_{\de/\ve}(s,u)$ is defined by 
\beq
\label{gdefve}
g_{\de/\ve}(s,u)=(1+u\de/\ve\gamma(s))^2\,,
\eeq
where $\ga(s)$ is   a function of $s$ and we  assume that $\ga\in C_0^\infty((-1,1))$ and $\|\ga\|_{L^\infty((-1,1))}<1$. We note that this implies that $g_{\defve}\in C^\infty(V)$ and that for all $0<\de\leqslant \ve\leqslant 1$,  the bounds 
\beq
\label{bounds}
0<\big(1-\|\ga\|_{L^\infty((-1,1))}\big)^2\leqslant\|g_{\de/\ve}\|_{L^\infty(V)}\leqslant\big(1+\|\ga\|_{L^\infty((-1,1))}\big)^2<4
\eeq
hold true. The waveguide is obtained by identifying the boundary of $E_{1,\de}$ corresponding to $s=0$ with the boundary of $V_{\de,\ve}$ corresponding to $s=-1$ and the boundary of $E_{2,\de}$ corresponding to $s=0$ with the boundary of $V_{\de,\ve}$ corresponding to $s=1$.

We denote by $\HH_{\de,\ve}$ the  Hilbert space 
\be 
\HH_{\de,\ve}:=
L^2(E_{1,\de})\oplus L^2(E_{2,\de})\oplus L^2(V_{\de,\ve})\,,
\ee
where $L^2(E_{j,\de})\equiv L^2(E,\hmet_\de)$, $j=1,2$, and $L^2(V_{\de,\ve})\equiv L^2(V,\gmet_\deve)$. Moreover we denote by a capital Greek letter a vector in $\HH_{\de,\ve}$, $\Psi=(\psi_1,\psi_2,\psi_v)$.  Given two vectors $\Phi\equiv(\phi_1,\phi_2,\phi_v)\in\HH_\deve$ and $\Psi\equiv(\psi_1,\psi_2,\psi_v)\in\HH_\deve$ the scalar product $(\Phi,\Psi)_{\HH_\deve}$ and the norm $\|\Psi\|_{\HH_\deve}$ read respectively
\be
(\Phi,\Psi)_{\HH_\deve}=
\sum_{k=1,2}\int_{0}^\infty\int_0^1 
\overline{\phi_k}\psi_k\det[\hmet_\de]^{1/2}dsdu
+\int_{-1}^1\int_0^1
\overline{\phi_v}\psi_v
\det[\gmet_{\de,\ve}]^{1/2}dsdu\,,
\ee
with 
\beq
\label{determ}
\det[\hmet_\de]=\de^2\;,\quad
\det[\gmet_{\de,\ve}]=\de^2\ve^2g_\defve\,,
\eeq
and $\|\Psi\|_{\HH_\deve}=(\Psi,\Psi)_{\HH_\deve}^{1/2}$.  In $\HH_{\de,\ve}$ we define the sesquilinear form  $Q_{\de,\ve}$ 
\be
\begin{aligned}
Q_{\deve}[\Phi,\Psi]:=&
\sum_{k=1,2}\int_{0}^\infty\int_0^1 
\bigg[
\overline{\pd{\phi_k}{s}}\pd{\psi_k}{s}
+\frac{1}{\de^2}
\overline{\pd{\phi_k}{u}}\pd{\psi_k}{u}\bigg]
\det[\hmet_\de]^{1/2}dsdu
\\
&
+\int_{-1}^1\int_0^1
\bigg[\frac{1}{\ve^2g_{\de/\ve}}
\overline{\pd{\phi_v}{s}}\pd{\psi_v}{s}+\frac{1}{\de^2}
\overline{\pd{\phi_v}{u}}\pd{\psi_v}{u}\bigg]
\det[\gmet_{\de,\ve}]^{1/2}dsdu\,.
\end{aligned}
\ee
Below we shall replace the term sesquilinear form   by quadratic form (this is justified by the polarization theorem, see, e.g., \cite{kato:80}) and we shall also use the notation $Q_\deve[\Psi]\equiv Q_\deve[\Psi,\Psi]$.  Let $\mathring{C}^{\infty}$ be the set 
\beq
\label{Cring}
\begin{aligned}
\mathring{C}^{\infty}:=\{&\Psi=(\psi_1,\psi_2,\psi_v)|\,
\psi_1,\psi_2\in C_0^\infty(\overline{E})\,,\;\psi_v \in C^\infty(V)\,;\\
& \psi_1(s,0)=\psi_1(s,1)=
\psi_2(s,0)=\psi_2(s,1)=\psi_v(s,0)=\psi_v(s,1)=0\,;\\
&\big[\partial_s^k\psi_1\big](0,u)=\big[(-\ve)^{-k}\partial_s^k\psi_v\big](-1,u)\,;\;
\big[\partial_s^k\psi_2\big](0,u)=\big[\ve^{-k}\partial_s^k\psi_v\big](1,u),\,\forall k \in \NN_0
\}\,,
\end{aligned}
\eeq
where we denoted by $\overline E$ the closure of $E$. The domain $D(Q_{\de,\ve})$  of the quadratic form $Q_{\de,\ve}$ is the closure of $\mathring{C}^\infty$ equipped  with the norm
\be
\|\Psi\|_{Q_{\de,\ve}}:=
\big(\|\Psi\|^2_{\HH_{\de,\ve}}+Q_{\de,\ve}[\Psi]\big)^{1/2}\,.
\ee

We denote by $H_{\de,\ve}$ the unique selfadjoint operator in $\HH_{\de,\ve}$ associated to the quadratic form $Q_{\de,\ve}$
\[
D( H_\deve):=
\{\Psi\in D(Q_{\de,\ve})|\;\forall\, \Phi\in D(Q_\deve)\,,
Q_{\de,\ve}[\Phi,\Psi]=
\langle\Phi,\Xi\rangle_{\HH_\deve}\,;\; \Xi\in \HH_\deve\}
\]
\[
H_\deve\Psi:=\Xi .
\]

\subsection{Unitarily equivalent Hamiltonian}
This section is devoted to the proof of Proposition \ref{p:2.1} below. The proposition allows us to rewrite the Hamiltonian $H_{\de,\ve}$ in terms of the unitarily equivalent Hamiltonian $\tilde H_{\de,\ve}$. The latter can be explicitly written in terms of a differential operator in the ``flat'' Hilbert space $\tilde \Hbb_\ve$. 

Let  $L^2(E)$ and $L^2(V)$  be the complex Hilbert spaces endowed with
the  norms 
\be
\|\tilde\psi_j\|_{L^2(E)}^2:=\int_{0}^\infty\int_0^1 |\tilde\psi_j|^2 dsdu\,,\;j=1,2\,;\quad
\|\tilde\psi_v\|_{L^2(V)}^2:=\int_{-1}^1\int_0^1 |\tilde\psi_j|^2 dsdu\,.
\ee

Let us denote by $\tilde\HH_\ve $ the complex Hilbert space 
\be
\tilde\HH_\ve
:=
L^2(E)\oplus L^2(E)\oplus L^2(V,\ve dsdu)\,.
\ee
Given two vectors $\tilde\Phi\equiv(\tilde\phi_1,\tilde\phi_2,\tilde\phi_v)\in\tilde\HH_\ve$ and $\tilde\Psi\equiv(\tilde\psi_1,\tilde\psi_2,\tilde\psi_v)\in\tilde\HH_\ve$ the scalar product $(\tilde\Phi,\tilde\Psi)_{\tilde\HH_\ve}$ and the norm $\|\tilde\Psi\|_{\tilde\HH_\ve}$ read
\be
(\tilde\Phi,\tilde\Psi)_{\tilde\HH_\ve}=
\sum_{k=1,2}\int_{0}^\infty\int_0^1 
\overline{\tilde\phi_k}\tilde\psi_kdsdu
+\ve \int_{-1}^1\int_0^1
\overline{\tilde\phi_v}\tilde\psi_v dsdu\,,
\ee
\begin{equation}\label{2.4a}
\|\tilde\Psi\|_{\tilde\HH_\ve}=\Big(\|\tilde\psi_1\|_{L^2(E)}^2+\|\tilde\psi_2\|_{L^2(E)}^2+\ve\|\tilde\psi_v\|_{L^2(V)}^2\Big)^{1/2}
\,.
\end{equation}
For all $0<\ve\leqslant 1$ we denote by $U_\deve$ the unitary map $U_\deve:\HH_\deve\to \tilde\HH_\ve$ defined by
\beq
\label{Uv}
\begin{aligned}
(\tilde\psi_1,\tilde\psi_2,\tilde\psi_v)\equiv
U_\deve (\psi_1,\psi_2,\psi_v):=&
(\det[\hmet_\de]^{1/4}\psi_1,\det[\hmet_\de]^{1/4}\psi_2,\ve^{-1/2}\det[\gmet_\deve]^{1/4}\psi_v)\\
=&
(\de^{1/2}\psi_1,\de^{1/2}\psi_2,\de^{1/2}g_\defve^{1/4}\psi_v)\,,
\end{aligned}
\eeq
where we used equation \eqref{determ}. 

For all $0<\de\leqslant \ve\leqslant 1$, we denote by $\Hunoring$ the closure of $\mathring{C}^\infty$, defined in equation \eqref{Cring}, with respect to the norm 
\beq
\label{H1ringnorm}
\|\Psi\|_{\Hunoring}^2:=
\|\Psi\|_{\tilde\HH_\ve}^2
+\sum_{j=1,2}\Big[\|\partial_s\psi_j\|_{L^2(E)}^2+\de^{-2}\|\partial_u\psi_j\|_{L^2(E)}^2\Big]
+\ve\Big[\ve^{-2}\|\partial_s\psi_v\|_{L^2(V)}^2+\de^{-2}\|\partial_u\psi_v\|_{L^2(V)}^2\Big],
\eeq
and by $\Hduering$ the closure of $\mathring{C}^\infty$ with respect to the norm
\[
\|\Psi\|_{\Hduering}^2:=
\|\Psi\|_{\Hunoring}^2+\sum_{j=1,2}\Big[\|\partial^2_s\psi_j\|_{L^2(E)}^2+\de^{-4}\|\partial^2_u\psi_j\|_{L^2(E)}^2\Big]+\ve\Big[\ve^{-4}\|\partial^2_s\psi_v\|_{L^2(V)}^2+\de^{-4}\|\partial^2_u\psi_v\|_{L^2(V)}^2\Big].
\]
We note that  $\Hunoring$ and  $ \Hduering$ coincide with 
\beq
\label{H1ring}
\begin{aligned}
 \Hunoring=\{
 \tilde\Psi\equiv(\tilde\psi_1,\tilde\psi_2,\tilde\psi_v)\in\tilde\HH_\ve|\,
 \|\tilde \Psi\|_{\Hunoring}<\infty\,;\;& \tilde\psi_1\big|_{u=0,1}=\tilde\psi_2\big|_{u=0,1}=0\,;\;
\tilde\psi_v\big|_{u=0,1}=0\,;\\
&\tilde\psi_1\big|_{s=0}=\tilde\psi_v\big|_{s=-1}\,,\;\tilde\psi_2\big|_{s=0}=\tilde\psi_v\big|_{s=1}
\}\,,
\end{aligned}
\eeq
\beq
\label{H2ring}
\begin{aligned}
\Hduering=\bigg\{
\tilde\Psi\equiv(\tilde\psi_1,\tilde\psi_2,\tilde\psi_v)\in\tilde\HH_\ve|\,\|\tilde \Psi\|_{\Hduering}<\infty\,;\;& \tilde\psi_1\big|_{u=0,1}=\tilde\psi_2\big|_{u=0,1}=0\,;\;
\tilde\psi_v\big|_{u=0,1}=0\,;\\
&\tilde\psi_1\big|_{s=0}=\tilde\psi_v\big|_{s=-1}\,;\;\tilde\psi_2\big|_{s=0}=\tilde\psi_v\big|_{s=1}\,;
\\
&\pd{\tilde\psi_1}{s}\bigg|_{s=0}=-\frac{1}{\ve}\pd{\tilde\psi_v}{s}\bigg|_{s=-1}\,;\;\pd{\tilde\psi_2}{s}\bigg|_{s=0}=\frac{1}{\ve}\pd{\tilde\psi_v}{s}\bigg|_{s=1}
\bigg\}\,.
\end{aligned}
\eeq
In formulae \eqref{H1ring} and \eqref{H2ring}, in the boundary values, the symbols $\big|_{u=0,1}$ and $\big|_{s=0,\pm1}$ are understood as trace operators. 
%

\begin{proposition}\label{p:2.1}
The Hamiltonian $H_\deve$ is unitarily equivalent to the Hamiltonian $\tilde H_{\deve}:D(\tilde H_\deve)\subset \tilde\HH_\ve\to\tilde \HH_\ve$ defined by
\beq
\label{DtHdeve}
D(\tilde H_\deve)=\Hduering
\eeq
\beq
\label{tHdeve}
\tilde H_\deve(\tilde\psi_1,\tilde\psi_2,\tilde\psi_v)\\
=
\bigg(\bigg[
-\pd{^2}{s^2}-
 \frac{1}{\de^2}\pd{^2}{u^2}\bigg]\tilde\psi_1,
 \bigg[
-\pd{^2}{s^2}-
 \frac{1}{\de^2}\pd{^2}{u^2}\bigg]\tilde\psi_2,
\frac{1}{\ve^2}\tilde{L}_{\defve}\tilde\psi_v
\bigg)\,,
\eeq
where $\tilde{L}_{\defve}$ denotes the differential operator
\beq
\label{L}
\tilde{L}_\defve:=-\frac{1}{g_\defve}\pd{^2}{s^2}
-\bigg[\pd{}{s}\frac{1}{g_\defve}\bigg]\pd{}{s}
+W_\defve
-\frac{1}{(\de/\ve)^2}\pd{^2}{u^2}
\eeq
with 
\beq
\label{W}
W_\defve=
-\frac{1}{4}\frac{\ga^2(s)}{(1+u\de/\ve\gamma(s))^2}
+\frac{1}{2}\frac{u\de/\ve\ddot\gamma(s)}{(1+u\de/\ve\gamma(s))^3}
-\frac{5}{4}\frac{\big(u\de/\ve\dot\gamma(s)\big)^2}{(1+u\de/\ve\gamma(s))^4}
\,.
\eeq
\end{proposition}
\begin{proof}
We prove first that the quadratic form $Q_\deve$ in $\HH_\deve$ is unitarily equivalent to the quadratic form $\tilde Q_\deve$ in $\tilde \HH_\ve$ defined by 
\be
\begin{aligned}
\tilde Q_{\deve}[\tilde\Phi,\tilde\Psi]:=&
\sum_{k=1,2}\int_{0}^\infty\int_0^1 
\bigg[
\overline{\pd{\tilde \phi_k}{s}}\pd{\tilde \psi_k}{s}
+\frac{1}{\de^2}
\overline{\pd{\tilde \phi_k}{u}}\pd{\tilde \psi_k}{u}\bigg]
dsdu
\\
&
+\ve\int_{-1}^1\int_0^1
\frac{1}{\ve^2}\bigg(\frac{1}{g_{\de/\ve}}
\overline{\pd{\tilde \phi_v}{s}}\pd{\tilde \psi_v}{s}+\widetilde W_{\defve}\tilde \phi_v\tilde \psi_v\bigg)
dsdu
\\
&
+\ve\int_{-1}^1\int_0^1
\frac{1}{\de^2}\bigg(
\overline{\pd{\tilde \phi_v}{u}}\pd{\tilde\psi_v}{u}
+\widetilde{\widetilde W}_{\defve}\tilde \phi_v\tilde \psi_v
\bigg)
dsdu
\end{aligned}
\ee
with $D(\tilde Q_\deve)=\Hunoring$ (see formula \eqref{H1ring} for the definition of $\Hunoring$), and 
\beq
\label{tildeW}
\widetilde W_\deve=
-\pd{}{s}\bigg[\frac{1}{g_{\de/\ve}^{3/4}}\pd{}{s}(g_\defve^{-1/4})\bigg]
+\frac{1}{g_{\de/\ve}^{1/2}}
\bigg(\pd{}{s}g_\defve^{-1/4}\bigg)^2\,,
\eeq
\[
\widetilde{\widetilde W}_\deve=
-\pd{}{u}\bigg[g_{\de/\ve}^{1/4}\pd{}{u}(g_\defve^{-1/4})\bigg]
+g_{\de/\ve}^{1/2}
\bigg(\pd{}{u}g_\defve^{-1/4}\bigg)^2\,.
\]
Let us denote by $\tilde Q_\deve^0$ the restriction of the quadratic form $\tilde Q_\deve$ to $ \mathring{C}^{\infty}$, see the definition \eqref{Cring}.  Since $g_\defve\in C^\infty(V)$ and the bounds \eqref{bounds} hold true, the norm 
\be
\|\tilde\Psi\|_{\tilde Q_{\de,\ve}}:=
\big(\|\tilde\Psi\|^2_{\tilde\HH_{\de,\ve}}+\tilde Q_{\de,\ve}[\tilde\Psi]\big)^{1/2}\,,
\ee
is equivalent to the $\Hunoring$-norm defined in equation \eqref{H1ringnorm}. Then the quadratic form $\tilde Q_{\de,\ve}$ coincides with the closure of $\tilde Q_{\de,\ve}^{0}$ in the norm $\|\cdot\|_{\tilde Q_{\de,\ve}}$. For any function $\tilde\Psi\in \mathring{C}^{\infty}$, one has $\Psi=U_\deve^{-1}\tilde \Psi \in \mathring{C}^{\infty}$, where $U_\deve$ is the unitary map defined in   \eqref{Uv} and $U_\deve^{-1}$ its inverse.

Let us define 
\be
\tilde Q^\#_{\deve}[\tilde \Phi,\tilde \Psi]:=Q_\deve[U_\deve^{-1}\tilde \Phi, U_\deve^{-1}\tilde \Psi_\deve]\,.
\ee
We shall prove that for any $\tilde \Psi, \tilde \Phi \in \mathring{C}^{\infty}$, one has that  $\tilde Q^\#_{\deve}[\tilde \Phi,\tilde \Psi]=\tilde Q_{\deve}[\tilde\Phi,\tilde\Psi]$. From the definitions of $U_\deve$  and $Q_\deve$ we have 
\be
\begin{aligned}
\tilde Q_{\deve}^\#[\tilde \Phi,\tilde \Psi]
=&\sum_{k=1,2}\int_{0}^\infty\int_0^1 
\bigg[
\overline{\pd{\tilde \phi_k}{s}}\pd{\tilde \psi_k}{s}
+\frac{1}{\de^2}
\overline{\pd{\tilde \phi_k}{u}}\pd{\tilde \psi_k}{u}\bigg]
dsdu
\\
&
+\int_{-1}^1\int_0^1
\bigg[\frac{1}{\ve^2g_{\de/\ve}^{1/2}}
\overline{\pd{}{s}(g_\defve^{-1/4}\tilde \phi_v)}\pd{}{s}(g_\defve^{-1/4} \tilde\psi_v)+\frac{g_{\de/\ve}^{1/2}}{\de^2}
\overline{\pd{}{u}(g_\defve^{-1/4}\tilde \phi_v)}\pd{}{u}(g_\defve^{-1/4} \tilde\psi_v)\bigg]
\ve dsdu\,.
\end{aligned}
\ee
We note that 
\be
\begin{aligned}
&\int_{-1}^1\int_0^1
\bigg[\frac{1}{\ve^2g_{\de/\ve}^{1/2}}
\overline{\pd{}{s}(g_\defve^{-1/4}\tilde \phi_v)}\pd{}{s}(g_\defve^{-1/4} \tilde\psi_v)\bigg]
\ve dsdu=
\\
=
&
\int_{-1}^1\int_0^1
\frac{1}{\ve^2g_{\de/\ve}}
\overline{\pd{\tilde \phi_v}{s}}\pd{\tilde\psi_v}{s}\ve dsdu\\
&+\int_{-1}^1\int_0^1
\bigg[\frac{1}{\ve^2g_{\de/\ve}^{3/4}}\pd{}{s}(g_\defve^{-1/4})
\bigg(\overline{\pd{\tilde \phi_v}{s}} \tilde\psi_v+\overline{ \tilde\phi_v}\pd{\tilde \psi_v}{s}\bigg)
+\frac{1}{\ve^2g_{\de/\ve}^{1/2}}
\bigg(\pd{}{s}g_\defve^{-1/4}\bigg)^2\tilde \phi_v \tilde\psi_v\bigg]\ve dsdu\,,
\end{aligned}
\ee
and 
\be
\begin{aligned}
&\int_{-1}^1\int_0^1
\frac{g_{\de/\ve}^{1/2}}{\de^2}
\overline{\pd{}{u}(g_\defve^{-1/4}\tilde \phi_v)}\pd{}{u}(g_\defve^{-1/4} \tilde\psi_v)
\ve dsdu
\\
=
&
\int_{-1}^1\int_0^1
\frac{1}{\de^2}
\overline{\pd{\tilde \phi_v}{u}}\pd{\tilde\psi_v}{u}\ve dsdu\\
&+\int_{-1}^1\int_0^1
\bigg[\frac{g_{\de/\ve}^{1/4}}{\de^2}\pd{}{u}(g_\defve^{-1/4})
\bigg(\overline{\pd{\tilde \phi_v}{u}} \tilde\psi_v+\overline{ \tilde\phi_v}\pd{\tilde \psi_v}{u}\bigg)
+\frac{g_{\de/\ve}^{1/2}}{\de^2}
\bigg(\pd{}{u}g_\defve^{-1/4}\bigg)^2\tilde \phi_v \tilde\psi_v\bigg]\ve dsdu\,.
\end{aligned}
\ee
By integration by parts in $s$ one has 
\be
\begin{aligned}
&\int_{-1}^1\int_0^1
\bigg[\frac{1}{\ve^2g_{\de/\ve}^{3/4}}\pd{}{s}(g_\defve^{-1/4})
\bigg(\overline{\pd{\tilde \phi_v}{s}} \tilde\psi_v+\overline{ \tilde\phi_v}\pd{\tilde \psi_v}{s}\bigg)
+\frac{1}{\ve^2g_{\de/\ve}^{1/2}}
\bigg(\pd{}{s}g_\defve^{-1/4}\bigg)^2\tilde \phi_v \tilde\psi_v\bigg]\ve
 dsdu\\
=&
\int_{-1}^1\int_0^1
\frac{1}{\ve^2}\widetilde W_\defve\tilde \phi_v \tilde\psi_v\ve dsdu\,,
\end{aligned}
\ee
where $\widetilde W_{\defve}$ was defined in equation \eqref{tildeW} and we used the fact that the boundary terms are null because $\gamma\in C^\infty_0((-1,1))$. 
By integration by parts in $u$ one has 
\be
\begin{aligned}
&
\int_{-1}^1\int_0^1
\bigg[\frac{g_{\de/\ve}^{1/4}}{\de^2}\pd{}{u}(g_\defve^{-1/4})
\bigg(\overline{\pd{\tilde \phi_v}{u}} \tilde\psi_v+\overline{ \tilde\phi_v}\pd{\tilde \psi_v}{u}\bigg)
+\frac{g_{\de/\ve}^{1/2}}{\de^2}
\bigg(\pd{}{u}g_\defve^{-1/4}\bigg)^2\tilde \phi_v \tilde\psi_v\bigg]\ve dsdu
\\
=&
\int_{-1}^1\int_0^1
\frac{1}{\de^2}\widetilde {\widetilde W}_\defve\tilde \phi_v \tilde\psi_v\ve dsdu\,,
\end{aligned}
\ee
where $\widetilde{\widetilde W}_{\defve}$ was defined in equation \eqref{tildeW} and we used the fact that the boundary terms are null because $\tilde \phi_v\big|_{u=0,1}= \tilde\psi_v\big|_{u=0,1}=0$. 

Let $\tilde H_\deve^\#$ the Hamiltonian in $\tilde \HH_\ve$ associated to the quadratic form $\tilde Q_\deve$. By integrating by parts in $s$ and $u$, one sees that for any $\tilde \Phi, \tilde \Psi \in \mathring{C}^{\infty}$
\beq
\label{changes}
\begin{aligned}
&\tilde Q_{\deve}[\tilde\Phi,\tilde\Psi]=
\sum_{k=1,2}\int_{0}^\infty\int_0^1 
\overline {\tilde \phi_k}\bigg[-
\pd{^2\tilde \psi_k}{s^2}
-\frac{1}{\de^2}\pd{^2\tilde \psi_k}{u^2}\bigg]
dsdu
\\
+&
\int_{-1}^1\int_0^1
\overline {\tilde \phi_v}\bigg[\frac{1}{\ve^2}\bigg(-\frac{1}{g_{\de/\ve}}
\pd{^2\tilde \psi_v}{s^2}-
\bigg(\pd{}{s}\frac{1}{g_{\de/\ve}}\bigg)
\tilde \psi_v+
\big(\widetilde W_{\defve}+(\ve/\de)^2\widetilde{\widetilde W}_{\defve}\big)\tilde \psi_v\bigg)-\frac{1}{\de^2}
\pd{^2\tilde\psi_v}{u^2}\bigg]\ve
dsdu
 \\
- &
\int_0^1\overline{\tilde\phi_1}\big|_{s=0}\bigg[\pd{\tilde\psi_1}{s}\bigg|_{s=0}+\frac{1}{\ve}\pd{\tilde\psi_v}{s}\bigg|_{s=-1}\bigg]
du -
\int_0^1\overline{\tilde\phi_2}\big|_{s=0}\bigg[\pd{\tilde\psi_2}{s}\bigg|_{s=0}
-
\frac{1}{\ve}\pd{\tilde\psi_v}{s}
\bigg|_{s=1}\bigg]du\,,
\end{aligned}
\eeq
where we used the fact that $g_{\defve}(\pm1,u)=1$ and the fact that $\tilde \Phi\in\Hunoring $. A straightforward calculation gives
\be
-\bigg(\pd{}{s}\frac{1}{g_{\de/\ve}}\bigg)+
\widetilde W_{\defve}+(\ve/\de)^2\widetilde{\widetilde W}_{\defve}=W_{\defve}\,.
\ee
Then by the first representation theorem, see, e.g., \cite{kato:80}, we have that the Hamiltonian $\tilde H_\deve^\#$ acts as $\tilde H_\deve$ defined in equation \eqref{tHdeve}.  Moreover the domain of $\tilde H_\deve^\#$ is given by the functions in $\Hunoring$ such that the boundary terms in equation  \eqref{changes} are zero and 
\be
\|\tilde H_\deve^\#\tilde\Psi\|_{\tilde\HH_\ve}<\infty\,.
\ee
It is easy to convince oneself that since $g_\defve\in C^\infty(V)$ and the bounds \eqref{bounds} hold true, one has $ D(\tilde H_\deve^\#)=\Hduering$ which implies $\tilde H^\#_\deve\equiv \tilde H_\deve$. 
\end{proof}

\section{Approximate solution of the resolvent equation}
\label{sec2}

In this section we give an approximate solution to the resolvent equation $(\tilde H_\deve-z)\tilde\Psi=\tilde\Xi$, for $z\in\CO\backslash\RE$ and  for some suitable choice of $\tilde\Xi \in  \tilde \HH_\ve$, see Theorem \ref{resconv} below. In the analysis we are forced to renormalize the spectral parameter $z$ to $z+n^2\pi^2/\de^2$, with $n$ integer. In the final part of the section we discuss the behavior of the approximate solution as $\ve\to 0$. 

 We denote by $h_v$ the  Hamiltonian in $L^2((-1,1))$ 
\beq
\label{Dhv}
D(h_v):=\{y\in H^2((-1,1))|\,y'(\pm1)=0\}
\eeq
\beq
\label{hv}
h_v:=-\frac{d^2}{ds^2}-\frac{\ga^2(s)}{4}\,.
\eeq
For $n=1,2,3,...$ we denote by  $y_{n}(s)$ the (real) eigenfunctions of $h_v$ and by $\la_n$ the corresponding eigenvalues arranged in increasing order
\be
h_vy_n=-y_n''-\frac{\ga^2}{4}y_n=\la_ny_n\;;\qquad
y_n'(\pm1)=0\,\qquad n\in \NN\;.
\ee
We assume the normalization $(y_n,y_m)_{L^2((-1,1))}=\delta_{n,m}$, $n,m\in\NN$. For any $z\in\CO\backslash \RE$ we denote by $r_v(z)$ the resolvent of $h_v$,
\be
r_{v}(z):=(h_v-z)^{-1}\;;\qquad z\in\CO\backslash\RE\,.
\ee
For $z\in\CO\backslash\RE $ the operator $r_v(z)$ is bounded by $1/|\Im z|$, and $r_v(z):L^2((-1,1))\to D(h_v)$. The integral kernel of  $r_v(z)$ can be written as 
\beq
\label{rvve}
r_{v}(z;s,s')=
\sum_n\frac{y_{n}(s)y_{n}(s')}{\la_{n}-z}\,.
\eeq
For $n$ large enough one has $(n-1/2)^2\pi^2/4<\la_n<(n+1/2)^2\pi^2/4$ and $\|y_n\|_{L^\infty((-1,1))}\leqslant c$ where $c$ does not depend on $n$ (see, e.g., \cite{Tit62}). Thus the series in \eqref{rvve} converges absolutely and pointwise for $s,s'\in[-1,1]$. As a function of $z$, the operator   $r_v(z)$ can be analytically continued to a linear bounded operator for $z\in\CO\backslash \{\la_n\}_{n\in\NA}$. 

We  give a formula for the kernel $r_v(z;s,s')$ which does not involve series (see, e.g., \cite[Ch 4.2]{Hel67}).  For any $z\in\CO\backslash \RE$,  let $\zeta_v(z)$ and $\eta_v(z)$ be two generic solutions of  the equations 
\beq
\label{rv4}
\begin{aligned}
&-\zeta_v''(z)+(-\ga^2/4-z)\zeta_v(z)=0\,;\quad&&\zeta_v'(z;-1)=0\\
&-\eta_v''(z)+(-\ga^2/4-z)\eta_v(z)=0\,;&&\eta_v'(z;1)=0
\end{aligned}
\eeq
and let $\WW_v(z;s)$ be the Wronskian 
\be
\WW_v(z;s):=\eta_v(z;s)\zeta_v'(z;s)-\zeta_v(z;s)\eta_v'(z;s)\,.
\ee
We note that $\WW_v(z;s)$ does not depend on $s$,
and we set $\WW_v(z)\equiv \WW_v(z;s)$. 
The integral kernel of $r_v(z)$ reads
\beq
\label{rv3}
r_v(z;s,s')=
\left\{\begin{aligned}
\frac{\zeta_v(z;s)\eta_v(z;s')}{\WW_v(z)}&\quad s\leqslant s'\\
\frac{\eta_v(z;s)\zeta_v(z;s')}{\WW_v(z)}&\quad s>s'
\end{aligned}
\right.
\eeq

We distinguish two cases:
\begin{itemize}
\item[\emph{Case 1.}] Zero  is not an eigenvalue of  the  Hamiltonian  $h_v$ defined in \eqref{Dhv} - \eqref{hv}
\item[\emph{Case 2.}]  Zero is an eigenvalue  of the Hamiltonian $h_v$ defined in \eqref{Dhv} - \eqref{hv}. Then we  denote by $n^*$ the integer corresponding to the eigenvalue zero, i.e., $\la_{n^*}=0$ and by  $y^*\equiv y_{n^*}$ the corresponding eigenfunction, i.e.,
\beq
\label{eqv*}
h_vy^* =  -\frac{d^2}{ds^2}y^*-\frac{\ga^2}{4}y^*=0
\eeq
with ${y^*}'(\pm1)=0$ and $\|y^*\|_{L^2((-1,1))}=1$. We define the constants 
\beq
\label{al1al2}
\al_1:=y^*(-1)\;;\quad \al_2:=y^*(1)\,.
\eeq
By the definition of $h_v$ we have that if $y_n$ is an eigenfuction of $h_v$ then so is $\bar y_n$. Hence we can assume that   $y^*$  is real, which in turns implies $\al_1,\al_2\in\RE$.
\end{itemize}

Let us denote by $\chi_n(u)$ the functions 
\beq
\label{chin}
\chi_n(u):=\sqrt{2}\sin(n\pi u)\;;\quad n\in\NN\,.
\eeq

\begin{definition}
\label{hatpsi12v}
For any vector $\tilde{\Xi}_n\equiv(f_1\chi_n,f_2\chi_n,0)$ with $f_1,f_2\in L^2((0,\infty))$ and $z\in \CO\backslash \RE$,  with $\Im\sqrt{z}>0$,  we denote by $\hatPsi$ the vector  $\hatPsi\equiv(\hatpsiuu,\hatpsidd,\hatpsiv)$, where the functions 
 $\hatpsiuu$, $\hatpsidd$ and $\hatpsiv$ are defined by 
\begin{align}
\label{psi1}
\hatpsiuu(s,u)&:=\big[\big(r_0(z)f_1\big)(s)+\quu e^{i\sqrt{z}s}\big]\chi_n(u)\\
\label{psi2}
\hatpsidd(s,u)&:=\big[\big(r_0(z)f_2\big)(s)+\qdd e^{i\sqrt{z}s}\big]\chi_n(u)\\
\label{psiv}
\hatpsiv(s,u)&:=\ve\big[\xiuu r_v(\ve^2 z;s,-1)+\xidd r_v(\ve ^2 z;s,1)\big]\chi_n(u)\,,
\end{align}
and where we set 
\beq
\label{r0z}
\big(r_0(z)f_j\big)(s):=\int_0^\infty\bigg(\frac{ie^{i\sqrt{z}|s-s'|}}{2\sqrt{z}}-\frac{ie^{i\sqrt{z}s}e^{i\sqrt{z}s'}}{2\sqrt{z}}\bigg)f_j(s')ds'\;;\quad\Im\sqrt{z}>0\,,\;j=1,2\,.
\eeq
The constants $\quu$, $\qdd$, $\xiuu$ and $\xidd$ are fixed by the relations
\beq
\label{hatxixi}
\begin{aligned}
&\xiuu:=(\puu+i\sqrt{z}\quu)\;;\quad \puu:=\big(r_0(z)f_1\big)'(0)\\
&\xidd:=(\pdd+i\sqrt{z}\qdd)\;;\quad \pdd:=\big(r_0(z)f_2\big)'(0)
\end{aligned}
\eeq
and 
\beq
\label{qxi}
\begin{pmatrix}
\quu\\ \\
\qdd
\end{pmatrix}
:=\begin{pmatrix}
\ve r_v(\ve^2z;-1,-1)&\ve r_v(\ve^2z;-1,1)\\ \\
\ve r_v(\ve^2 z;1,-1)&\ve r_v(\ve^2 z;1,1)
\end{pmatrix}
\begin{pmatrix}
\xiuu\\ \\
\xidd
\end{pmatrix}
\eeq
\end{definition}

Our main result is expressed in the following  Theorem \ref{resconv} and  Theorem \ref{asymbeh},  the proofs of which are postponed to Section \ref{secth2} and Section \ref{secth1} respectively.

\begin{theorem}
\label{resconv}
For any vector $\tilde{\Xi}_n\equiv(f_1\chi_n,f_2\chi_n,0)$ with $f_1,f_2\in L^2((0,\infty))$  let us take $\hatPsi$ as   in Def. \ref{hatpsi12v}. Then $\hatPsi\in D(\tilde H_\deve)$, moreover for all $z\in\CO\backslash\RE$ there exists $\ve_0>0$ such that for all $0<\ve<\ve_0$ and for all $0<\de\leqslant\ve$ the following estimates hold true: 
\begin{itemize}
\item[\emph{Case 1.}]
\be
\bigg\|\bigg[\tilde H_\deve-\frac{n^2\pi^2}{\de^2}-z\bigg]\hatPsi-\tilde{\Xi}_n\bigg\|_{\tilde\HH_\ve}
\leqslant c\frac{\de}{\ve^{3/2}} \|\tilde{\Xi}_n\|_{\tilde\HH_\ve} \,;
\ee
\item[\emph{Case 2.}]
\be
\bigg\|\bigg[\tilde H_\deve-\frac{n^2\pi^2}{\de^2}-z\bigg]\hatPsi-\tilde{\Xi}_n\bigg\|_{\tilde\HH_\ve}\leqslant
c \frac{\de}{\ve^{5/2}} \|\tilde{\Xi}_n\|_{\tilde\HH_\ve} \,;
\ee
\end{itemize}
where $c$ is a  constant which does not depend on $\ve$, $f_1$, $f_2$ and $n$.
\end{theorem}

\begin{remark}
\label{remark}
We note  the following estimate for the resolvent of $\tilde H_\deve$:  if $\tilde{\Xi}_n\equiv(f_1\chi_n,f_2\chi_n,0)\in \tilde \HH_\ve$, then
\be
\bigg|\bigg(\tilde{\Xi}_n, \hatPsi-\bigg[\tilde H_\deve-\frac{n^2\pi^2}{\de^2}-z\bigg]^{-1}\tilde{\Xi}_n\bigg)_{\tilde\HH_\ve}\bigg|\leqslant
\frac{1}{|\Im z|}\big\|\tilde{\Xi}_n\big\|_{\tilde\HH_\ve}
\bigg\|\bigg[\tilde H_\deve-\frac{n^2\pi^2}{\de^2}-z\bigg]\hatPsi-\tilde{\Xi}_n\bigg\|_{\tilde\HH_\ve}\,.
\ee
\end{remark}

\begin{theorem}[Asymptotic behavior of the solution in the edges] 
\label{asymbeh}
Let us take  $\quu$, $\qdd$, $\xiuu$ and $\xidd$  as it was done in Eqs. \eqref{hatxixi} - \eqref{qxi}. Then:
\begin{itemize}
\item[\emph{Case 1.}]
\beq
\label{never}
\begin{pmatrix}
\quu\\ \\
\qdd
\end{pmatrix}
=\OO(\ve)
\begin{pmatrix}
p_1\\ \\
p_2
\end{pmatrix}
\;;\quad
\begin{pmatrix}
\xiuu\\ \\
\xidd
\end{pmatrix}
=\big[1+\OO(\ve)\big]
\begin{pmatrix}
p_1\\ \\
p_2
\end{pmatrix}
\,.
\eeq
\item[\emph{Case 2.}]
\beq
\label{q1q2asym2}
\begin{pmatrix}
\quu\\ \\
\qdd
\end{pmatrix}
=
\bigg[\frac{i\La_0}{\sqrt{z}}+\OO(\ve)\bigg]
\begin{pmatrix}
p_1\\ \\
p_2
\end{pmatrix}
\,;
\eeq
\beq
\label{xi1xi2asym2}
\begin{pmatrix}
\xiuu\\ \\
\xidd
\end{pmatrix}
=\bigg[\La_0^\perp+\OO(\ve)\bigg]
\begin{pmatrix}
p_1\\ \\
p_2
\end{pmatrix}
\eeq
where we denoted by $\La_0$ the projector
\beq
\label{proj}
\La_0=
\frac{1}{\al_1^2+\al_2^2}
\begin{pmatrix}
\al_1^2&\al_1\al_2\\ \\
\al_1\al_2&\al_2^2
\end{pmatrix}\quad\textrm{and}\qquad
\La_0^\perp=1-\La_0\,.
\eeq
\end{itemize}
\end{theorem}

Here and in the following for all $a\geqslant0$ we denote by $\OO(\ve^a)$ a $2\times 2$ matrix such that $\|\OO(\ve^a)\|_{\BB(\CO^2)}\leqslant c\ve^a$.

\section{Asymptotic behavior of the solution in the edges (proof of Theorem \ref{asymbeh})}
\label{secth1}

We devote this section to the proof of Theorem \ref{asymbeh}. We start with the proof of the following proposition: 

\begin{proposition}
\label{suprv}
For any $z\in\CO\backslash\RE$ there exists $\ve_0>0$ such that for all $0<\ve<\ve_0$ the following estimates hold true:
\begin{itemize}
\item[\emph{Case 1.}]
\beq
\label{estrv1}
  \sup_{s,s'\in[-1,1]}\big[|r_v(\ve^2z;s,s')|\big]\leqslant c\,;
\eeq
\item[\emph{Case 2.}]
\beq
\label{estrv2}
 \sup_{s,s'\in[-1,1]} \bigg[\bigg|r_v(\ve^2z;s,s')+\frac{y^*(s)y^*(s')}{\ve^2z}\bigg|\bigg]\leqslant c\,.
\eeq
where $c$ is a constant which does not depend on $\ve$.
\end{itemize}
\end{proposition}
\begin{proof}
We note that in \emph{Case 1} the series $
\sum_{n}\frac{1}{|\la_{n}|}$ is convergent and we use the formula \eqref{rvve} for the integral kernel $r_v(z;s,s')$. Then, in \emph{Case 1},
\be
|r_v(\ve^2 z;s,s')|=
\bigg|\sum_n\frac{y_{n}(s)y_{n}(s')}{\la_{n}-\ve^2z}\bigg|\leqslant
\sum_n\frac{|y_{n}(s)||y_{n}(s')|}{|\la_{n}-\ve^2z|}\,.
\ee
Then we use the fact that $\|y_{n}\|_{L^\infty((-1,1))}\leqslant c$ where $c$ does not depend on $n$, see, e.g., \cite{Tit62}, and the fact that for $\ve$ small enough  $|\la_{n}-\ve^2z|>2|\la_n|$. We obtain 
\be
|r_v(\ve^2 z;s,s')|
\leqslant
\max_n\big[\|y_n\|_{L^\infty((-1,1))}^2\big]
\sum_n\frac{1}{2|\la_{n}|}\leqslant c.
\ee
To prove the estimate \eqref{estrv2} we use again the formula \eqref{rvve}; we write the integral kernel $r_v(\ve^2 z;s,s')$ as
\be
r_v(\ve^2 z;s,s')=
-\frac{y^*(s)y^*(s')}{\ve^2z}
+\sum_{n\neq n^*}\frac{y_{n}(s)y_{n}(s')}{\la_{n}-\ve^2z}\,.
\ee
We recall that we denoted by $n^*$ the integer associated to the eigenvalue zero $\la_{n^*}=0$. The estimate \eqref{estrv2} follows from the fact that the series $\sum_{n\neq n^*}\frac{1}{|\la_{n}|}$ is convergent and by the same argument used for the analysis of \emph{Case 1}. 
\end{proof}

\begin{remark}
\label{l2normrv}
We note that Prop. \ref{suprv} implies that there exists $\ve_0$ such that for all $0<\ve<\ve_0:$ 
\begin{itemize}
\item[\emph{Case 1.}]
\be
\|r_v(\ve^2z;\cdot,\pm1)\|_{L^2((-1,1))} \leqslant 2\bigg( \sup_{s,s'\in[-1,1]}\big[|r_v(\ve^2z;s,s')|\big]\bigg)\leqslant c\,;
\ee
\item[\emph{Case 2.}]
\be
\bigg\|r_v(\ve^2z;\cdot,\pm1)+\frac{y^*y^*(\pm1)}{\ve^2z}\bigg\|_{L^2((-1,1))}
\leqslant 2\bigg(
 \sup_{s,s'\in[-1,1]} \bigg[\bigg|r_v(\ve^2z;s,s')+\frac{y^*(s)y^*(s')}{\ve^2z}\bigg|\bigg]\bigg)\leqslant c\,.
\ee
where $c$ is a constant which does not depend on $\ve$.
\end{itemize}
\end{remark}

We are now ready to give the proof of Theorem \ref{asymbeh}:
\begin{proof}[{\bf Proof of Theorem \ref{asymbeh}}]
We use the equalities  \eqref{hatxixi} in equation  \eqref{qxi} and obtain 
\be
\begin{pmatrix}
\quu\\ \\
\qdd
\end{pmatrix}
=
\begin{pmatrix}
\ve r_v(\ve^2z;-1,-1)&\ve r_v(\ve^2z;-1,1)\\ \\
\ve r_v(\ve^2 z;1,-1)&\ve r_v(\ve^2 z;1,1)
\end{pmatrix}
\begin{pmatrix}
p_1+i\sqrt{z}\quu\\ \\
p_2+i\sqrt{z}\qdd
\end{pmatrix}
\,.
\ee
We denote by $\La_\ve$ the matrix 
\be
\La_\ve:=
\begin{pmatrix}
r_v(\ve^2z;-1,-1)&r_v(\ve^2z;-1,1)\\ \\
r_v(\ve^2 z;1,-1)&r_v(\ve^2 z;1,1)
\end{pmatrix}\,,
\ee
and  we get the identity
\be
\begin{pmatrix}
\quu\\ \\
\qdd
\end{pmatrix}
=(\II-i\ve\sqrt{z}\La_\ve)^{-1}\ve\La_\ve
\begin{pmatrix}
p_1\\ \\
p_2
\end{pmatrix}\,.
\ee
In what follows we shall prove that $\II-i\ve\sqrt{z}\La_\ve$ is indeed invertible for $\ve$ small enough. 

In \emph{Case 1}, the estimate \eqref{estrv1} gives
\be
|r_v(\ve^2z; (-1)^j,(-1)^k)|\leqslant c\qquad j,k=1,2
\ee
which implies $\La_\ve=\OO(1)$ and,  consequently, that $\II-i\ve\sqrt{z}\La_\ve$ is invertible for small enough $\ve$ and that  the first estimate in equation \eqref{never} holds true. The second  estimate in the same equation follows directly from the definition of  $\xiuu$ and $\xidd$, see formula \eqref{hatxixi}.

In \emph{Case 2} we use the formula
\be
r_v(\ve^2z;(-1)^j ,(-1)^k)=-\frac{\al_j\al_k}{\ve^2z}+\sum_{n\neq n^*}
\frac{y_{n}((-1)^j)y_{n}((-1)^k)}{\la_{n}-\ve^2 z}\qquad j,k=1,2\,,
\ee
see the estimate \eqref{estrv2} in Proposition \ref{suprv}. Reminding that $y^*((-1)^k)=\al_k$, for $k=1,2$, it follows that 
\be
\ve \La_\ve=-\frac{1}{\ve z}
\begin{pmatrix}
\al_1^2&\al_1\al_2\\ \\
\al_1\al_2&\al_2^2
\end{pmatrix}
+
\ve \begin{pmatrix}
\sum_{n\neq n^*}
\frac{y_{n}(-1)y_{n}(-1)}{\la_{n}-\ve^2 z}&\sum_{n\neq n^*}
\frac{y_{n}(-1)y_{n}(1)}{\la_{n}-\ve^2 z}\\ \\
\sum_{n\neq n^*}
\frac{y_{n}(1)y_{n}(-1)}{\la_{n}-\ve^2 z}&\sum_{n\neq n^*}
\frac{y_{n}(1)y_{n}(1)}{\la_{n}-\ve^2 z}
\end{pmatrix}
=-\frac{\al_1^2+\al_2^2}{\ve z}\La_0+\ve\widetilde\La_1,
\ee
where the matrix $\widetilde\La_1=\OO(1)$ (see Proposition \ref{suprv}).  To prove that  $\II-i\ve\sqrt{z}\La_\ve$ is  invertible for small enough $\ve$,  we rewrite 
\[\II-i\ve\sqrt{z}\La_\ve = \frac{c_0}{\ve} \Lambda_0 +\II + \ve \Lambda_1\]
with $c_0 = i (\al_1^2+\al_2^2) / \sqrt z $ and $\Lambda_1 = -i\sqrt z \widetilde \Lambda_1$. For any vector $\vv\in \CO^2$ we have that 
\[\begin{aligned}
\left\| \left(\frac{c_0}{\ve} \Lambda_0 +\II + \ve \Lambda_1\right) \vv \right\|_{\CO^2}^2    & =  \left\| \left(\frac{c_0}{\ve}+1\right)  \Lambda_0 \vv + \Lambda_0^\perp \vv + \ve \Lambda_1 \vv \right\|_{\CO^2}^2     \\ 
& \geq  \frac12 \left\| \left(\frac{c_0}{\ve}+1\right)  \Lambda_0 \vv  + \Lambda_0^\perp \vv\right\|_{\CO^2}^2  - \ve^2 \left\|  \Lambda_1 \vv \right\|_{\CO^2}^2 
  \\ 
& =  \frac12\left(\left|\frac{c_0}{\ve}+1\right|^2 \left\|  \Lambda_0 \vv  \right\|_{\CO^2}^2+ \|\Lambda_0^\perp \vv\|_{\CO^2}^2  \right)- \ve^2 \left\|  \Lambda_1 \vv \right\|_{\CO^2}^2  \\ 
&  \geq  \frac12 \left\|\vv  \right\|_{\CO^2}^2 - \ve^2 \left\|  \Lambda_1 \vv \right\|_{\CO^2}^2  \geq c \|\vv\|_{\CO^2}\,,
\end{aligned}\]
where we used the trivial inequality $\|\vv_1+\vv_2\|_{\CO^2}^2 \geq \|\vv_1\|_{\CO^2}^2/2 - \|\vv_2\|_{\CO^2}^2$; the  fact that $\left|\frac{c_0}{\ve}+1\right|^2 \geq \frac{(\Im c_0)^2}{\ve^2} \geq 1$ for small enough $\ve$; and the identity $ \left\|  \Lambda_0 \vv  \right\|_{\CO^2}^2+ \|\Lambda_0^\perp \vv\|_{\CO^2}^2 =  \left\| \vv  \right\|_{\CO^2}^2$. This proves that $\II-i\ve\sqrt{z}\La_\ve$ is invertible and that  
\begin{equation}\label{pj}(\II-i\ve\sqrt{z}\La_\ve)^{-1} = \OO(1)\,.\end{equation}
We get the following formula for the operator $(\II-i\ve\sqrt{z}\La_\ve)^{-1}\ve\La_\ve$
\be
(\II-i\ve\sqrt{z}\La_\ve)^{-1}\ve\La_\ve
=\bigg[i\frac{\al_1^2+\al_2^2}{\sqrt{z}}\La_0+\ve-i\ve^2\sqrt{z}\widetilde\La_1\bigg]^{-1}\bigg[-\frac{\al_1^2+\al_2^2}{z}\La_0+\ve^2\widetilde\La_1\bigg]\,.
\ee
Since, by Eq. \eqref{pj}, one has  
\begin{equation*}
\bigg[i\frac{\al_1^2+\al_2^2}{\sqrt{z}}\La_0+\ve-i\ve^2\sqrt{z}\widetilde\La_1\bigg]^{-1} = \OO(1/\ve),
\end{equation*}
it follows that 
\begin{equation}\label{age1}
\ve^2 \bigg[i\frac{\al_1^2+\al_2^2}{\sqrt{z}}\La_0+\ve-i\ve^2\sqrt{z}\widetilde\La_1\bigg]^{-1}\widetilde\La_1 = \OO(\ve).
\end{equation}Hence, 
\begin{equation}\label{age4}
(\II-i\ve\sqrt{z}\La_\ve)^{-1}\ve\La_\ve
=-\frac{\al_1^2+\al_2^2}{z}\bigg[i\frac{\al_1^2+\al_2^2}{\sqrt{z}}\La_0+\ve-i\ve^2\sqrt{z}\widetilde\La_1\bigg]^{-1}\La_0+\OO(\ve)\,.
\end{equation}
Next we use the identity
\begin{equation}\label{age2}\begin{aligned}
&\bigg[i\frac{\al_1^2+\al_2^2}{\sqrt{z}}\La_0+\ve-i\ve^2\sqrt{z}\widetilde\La_1\bigg]^{-1}\La_0  \\ 
=& 
\bigg[i\frac{\al_1^2+\al_2^2}{\sqrt{z}}\La_0+\ve\bigg]^{-1}\La_0+ i\ve^2\sqrt{z}  \bigg[i\frac{\al_1^2+\al_2^2}{\sqrt{z}}\La_0+\ve-i\ve^2\sqrt{z}\widetilde\La_1\bigg]^{-1} \widetilde\La_1 \bigg[i\frac{\al_1^2+\al_2^2}{\sqrt{z}}\La_0+\ve\bigg]^{-1}\La_0 . 
\end{aligned}\end{equation}
Since $\Lambda_0$ is a projection, one has
\begin{equation}\label{age3} \bigg[i\frac{\al_1^2+\al_2^2}{\sqrt{z}}\La_0+\ve\bigg]^{-1}\La_0 =  \bigg[i\frac{\al_1^2+\al_2^2}{\sqrt{z}}+\ve\bigg]^{-1}\La_0 =  -i\frac{\sqrt{z}}{\al_1^2+\al_2^2}\La_0  + \OO(\ve). \end{equation}
 Using identities  \eqref{age1} and \eqref{age3} in \eqref{age2} we get 
\begin{equation*}\bigg[i\frac{\al_1^2+\al_2^2}{\sqrt{z}}\La_0+\ve-i\ve^2\sqrt{z}\widetilde\La_1\bigg]^{-1}\La_0 =   -i\frac{\sqrt{z}}{\al_1^2+\al_2^2}\La_0  + \OO(\ve) ,
\end{equation*}
 which, together with \eqref{age4}, gives 
 \begin{equation*}
(\II-i\ve\sqrt{z}\La_\ve)^{-1}\ve\La_\ve
=\frac{i}{\sqrt{z}}\La_0  + \OO(\ve) 
\end{equation*}
and concludes the proof of \eqref{q1q2asym2}. 

The estimate \eqref{xi1xi2asym2} comes from the definition of $\xiuu$ and $\xidd$, see equation \eqref{hatxixi}, and from  equation \eqref{q1q2asym2}.
\end{proof}

\section{Asymptotic behavior of the solution in the vertex region (proof of Theorem \ref{resconv})}
\label{secth2}

We devote this section to the proof of Theorem \ref{resconv}. We first give some preliminary estimates on the function  $\hatpsiv$, see Proposition \ref{prop-estpsiv} below.

For any $z\in\CO\backslash\RE$, we denote by $r_v^{(0)}(z)$ the resolvent of the Neumann  Laplacian in $L^2((-1,1))$.    The integral kernel of $r_v^{(0)}(z)$ can be derived from formulas \eqref{rv4} - \eqref{rv3} by setting $\ga=0$; a straightforward calculation gives 
\beq
\label{rv03}
r_v^{(0)}(z;s,s')=
\left\{\begin{aligned}
-\frac{\cos(\sqrt{z}(s+1))\cos(\sqrt{z}(s'-1))}{\sqrt{z}\sin(2\sqrt{z})}&\quad s\leqslant s'\\
-\frac{\cos(\sqrt{z}(s-1))\cos(\sqrt{z}(s'+1))}{\sqrt{z}\sin(2\sqrt{z})}&\quad s>s'
\end{aligned}
\right.
\eeq

\begin{proposition}
For any $z\in\CO\backslash\RE$ there exists $\ve_0>0$ such that for all $0<\ve<\ve_0$ the following estimates hold true:
\beq
\label{jazz}
\bigg\|\frac{d}{ds}r_v^{(0)}(\ve^2z;\cdot,-1)\bigg\|_{L^2((-1,1))}\leqslant c \;;\quad
\bigg\|\frac{d}{ds}r_v^{(0)}(\ve^2z;\cdot,1)\bigg\|_{L^2((-1,1))}\leqslant c\,;
\eeq
where $c$ is a constant which does not depend on $\ve$. Moreover for any $g\in L^2((-1,1))$
\beq
\label{ddsrv0}
  \bigg\|\frac{d}{ds}r_v^{(0)}(\ve^2z)g\bigg\|_{L^2((-1,1))}\leqslant c\|g\|_{L^2((-1,1))}\,,
\eeq
where $c$ is a constant which does not depend on $\ve$.
\end{proposition}
\begin{proof}
By a direct computation
\beq
\label{onda}
\bigg\|\frac{d}{ds}r_v^{(0)}(\ve^2z;\cdot,\pm1)\bigg\|_{L^2((-1,1))}^2
=\int_{-1}^1\bigg|\frac{d}{ds}\frac{\cos(\sqrt{\ve^2z}(s\pm1))}{\sqrt{\ve^2z}\sin(2\sqrt{\ve^2z})}\bigg|^2ds=
\frac{\|\sin(\sqrt{\ve^2z}(\cdot\pm1))\|_{L^2((-1,1))}^2}{|\sin(2\sqrt{\ve^2z})|^2}
\eeq
The norms $\|\sin(\sqrt{\ve^2z}(\cdot-1))\|_{L^2((-1,1))}$ and $\|\sin(\sqrt{\ve^2z}(\cdot+1))\|_{L^2((-1,1))}$ can be bounded by 
\beq
\label{ondaonda1}
\|\sin(\sqrt{\ve^2z}(\cdot-1))\|_{L^2((-1,1))}^2=
\int_{-1}^1\bigg|\frac{e^{i\sqrt{\ve^2z}(s-1)}-e^{-i\sqrt{\ve^2z}(s-1)}}{2}\bigg|^2ds
\leqslant c\ve^2\,.
\eeq
and  similarly 
\beq
\label{ondaonda2}
  \|\sin(\sqrt{\ve^2z}(\cdot+1))\|_{L^2((-1,1))}^2\leqslant c\ve^2.
\eeq  
Using the latter estimates in equations \eqref{onda} we get 
\be
\bigg\|\frac{d}{ds}r_v^{(0)}(\ve^2z;\cdot,\pm1)\bigg\|_{L^2((-1,1))}^2\leqslant\frac{c\ve^2}{|\sin(2\sqrt{\ve^2z})|^2}
\,,
\ee
which, for $\ve$ small enough, imply  the bounds \eqref{jazz}, here we used the trivial inequality $|\sin w|^2 \geq c |w|^2$ which holds true for any $w\in \CO$ small enough. Also the estimate \eqref{ddsrv0} can be obtained through a direct calculation. For any  $g\in L^2((-1,1),ds)$
\be
\begin{aligned}
\frac{d}{ds}\big(r_v^{(0)}(z)g\big)(s)
=&
\frac{\sin(\sqrt{z}(s+1))}{\sin(2\sqrt{z})}\int_{s}^1\cos(\sqrt{z}(s'-1))g(s')
ds'\\
&+\frac{\sin(\sqrt{z}(s-1))}{\sin(2\sqrt{z})}
\int_{-1}^s\cos(\sqrt{z}(s'+1))g(s')ds'\,.
\end{aligned}
\ee
Then
\beq
\label{wanda}
\begin{aligned}
\bigg\|\frac{d}{ds}r_v^{(0)}(\ve^2z)g\bigg\|_{L^2((-1,1))}
\leqslant&
\bigg\|
\frac{\sin(\sqrt{\ve^2z}(\cdot+1))}{\sin(2\sqrt{\ve^2z})}\int_{\cdot}^1\cos(\sqrt{\ve^2z}(s'-1))g(s')
ds'\bigg\|_{L^2((-1,1))}\\
&+\bigg\|\frac{\sin(\sqrt{\ve^2z}(\cdot-1))}{\sin(2\sqrt{\ve^2z})}
\int_{-1}^\cdot\cos(\sqrt{\ve^2z}(s'+1))g(s')ds'\bigg\|_{L^2((-1,1))}\,.
\end{aligned}
\eeq
From the Cauchy-Schwarz inequality, for all $s\in(-1,1)$,
\be
\bigg|\int_{s}^1\cos(\sqrt{\ve^2z}(s'-1))g(s')ds'\bigg|
\leqslant
c\|g\|_{L^2((-1,1))}\,,
\ee
which implies 
\[
\begin{aligned}
& \bigg\|
\frac{\sin(\sqrt{\ve^2z}(\cdot+1))}{\sin(2\sqrt{\ve^2z})}\int_{\cdot}^1\cos(\sqrt{\ve^2z}(s'-1))g(s')
ds'\bigg\|_{L^2((-1,1))} \\ 
\leqslant & c
\frac{1}{|\sin(2\sqrt{\ve^2z})|}\|\sin(\sqrt{\ve^2z}(\cdot+1))\|_{L^2((-1,1))} \|g\|_{L^2((-1,1))}\,.
\end{aligned}
\]
Using again the estimates \eqref{ondaonda1} and \eqref{ondaonda2} we obtain 
\be
\bigg\|
\frac{\sin(\sqrt{\ve^2z}(\cdot+1))}{\sin(2\sqrt{\ve^2z})}\int_{\cdot}^1\cos(\sqrt{\ve^2z}(s'-1))g(s')
ds'\bigg\|_{L^2((-1,1))}\leqslant c\|g\|_{L^2((-1,1))}.
\ee
The second term at the r.h.s. of equation \eqref{wanda} is similar. Then bound \eqref{ddsrv0} follows from equation \eqref{wanda}.
\end{proof}

\begin{proposition}
\label{estimatesrv}
For any $z\in\CO\backslash\RE$ there exists $\ve_0>0$ such that for all $0<\ve<\ve_0$ the following estimates hold true:
\begin{itemize}
\item[\emph{Case 1.}]
\beq
\label{estddsrv1}
  \bigg\|\frac{d}{ds}r_v(\ve^2z;\cdot,-1)\bigg\|_{L^2((-1,1))}\leqslant c\;,\quad
  \bigg\|\frac{d}{ds}r_v(\ve^2z;\cdot,1)\bigg\|_{L^2((-1,1))}\leqslant c\,;
\eeq
\item[\emph{Case 2.}]
\beq
\label{estddsrv2}
  \bigg\|\frac{d}{ds}r_v(\ve^2z;\cdot,-1)+\frac{{y^*}'y^*(-1)}{\ve^2z}\bigg\|_{L^2((-1,1))}\leqslant c\;,\quad
  \bigg\|\frac{d}{ds}r_v(\ve^2z;\cdot,1)+\frac{{y^*}'y^*(-1)}{\ve^2z}\bigg\|_{L^2((-1,1))}\leqslant c\,.
\eeq
\end{itemize}
where $c$ is a constant which does not depend on $\ve$.
\end{proposition}
\begin{proof}
To prove the estimate \eqref{estddsrv1} we use the  well known  resolvent identity
\beq
\label{raglan}
\begin{aligned}
r_v(\ve^2 z;s,-1)=&\big[r_v(\ve^2 z)\de(\cdot+1)\big](s)=
\big[\big(r_v^{(0)}(\ve^2z)-r_v^{(0)}(\ve^2z)(-\ga^2/4)r_v(\ve^2 z)\big)\de(\cdot+1)\big]\\
=&r_v^{(0)}(\ve^2 z;s,-1)-\big[r_v^{(0)}(\ve^2z)(-\ga^2/4)r_v(\ve^2 z;\cdot,-1)\big](s)\,.
\end{aligned}
\eeq
Taking the derivative of the latter expression we obtain 
\be
 \begin{aligned}
  \bigg\|\frac{d}{ds}r_v(\ve^2z;\cdot,-1)\bigg\|_{L^2((-1,1))}=&
  \bigg\|\frac{d}{ds}r_v^{(0)}(\ve^2 z;\cdot,-1)-\frac{d}{ds}\big[r_v^{(0)}(\ve^2z)(-\ga^2/4)r_v(\ve^2 z;\cdot,-1)\bigg\|_{L^2((-1,1))}
  \\
  \leqslant&c[1+\|r_v(\ve^2 z;\cdot,-1)\|_{L^2((-1,1))}]\,,
  \end{aligned}
\ee
where we used the triangle inequality, estimates \eqref{jazz} and \eqref{ddsrv0}, and the fact that $\ga$ is bounded. As we have proven that   in \emph{Case 1}, $\|r_v(\ve^2 z;\cdot,-1)\|_{L^2((-1,1))}\leqslant c$, see Remark \ref{l2normrv}, we get the first estimate in the equation  \eqref{estddsrv1}, the proof of the second one is similar and we omit it. 

To prove the  estimates \eqref{estddsrv2}, we rewrite the equation \eqref{raglan} as 
\beq
\label{225}
\begin{aligned}
r_v(\ve^2 z;s,-1)
=
&
r_v^{(0)}(\ve^2 z;s,-1)-
\bigg[r_v^{(0)}(\ve^2z)(-\ga^2/4)\bigg(r_v(\ve^2 z;\cdot,-1)+\frac{y^*y^*(-1)}{\ve^2z}\bigg)\bigg](s)\\
&+ 
\bigg[r_v^{(0)}(\ve^2z)(-\ga^2/4)\frac{y^*y^*(-1)}{\ve^2z}\bigg](s)
\,.
\end{aligned}
\eeq
Then we use the fact that $y^*$ is the  solution of the problem  \eqref{eqv*} and we get 
\be
  \bigg[r_v^{(0)}(\ve^2z)(-\ga^2/4)\frac{y^*y^*(-1)}{\ve^2z}\bigg](s)=
  \bigg[r_v^{(0)}(\ve^2z)\frac{d^2}{ds^2}\frac{y^*y^*(-1)}{\ve^2z}\bigg](s)\,.
\ee
But
\be
  \begin{aligned}
   \bigg[r_v^{(0)}(\ve^2z)\frac{d^2}{ds^2}y^*\bigg](s)=&
 \int_{-1}^1r_v^{(0)}(\ve^2z;s,s')\frac{d^2}{d{s'}^2}y^*(s')ds'=
 -\int_{-1}^1 \frac{d}{ds'} r_v^{(0)}(\ve^2z;s,s')\frac{d}{ds'}y^*(s')ds'
 \\
 =&\int_{-1}^1 \frac{d^2}{d{s'}^2} r_v^{(0)}(\ve^2z;s,s')y^*(s')ds'=
    -\ve^2z[r_v^{(0)}(\ve^2z)y^*](s)-y^*(s)\,,
  \end{aligned}
\ee
where we used the fact that the boundary terms are null because 
\be
\frac{d}{ds}y^*\bigg|_{s=\pm1}=0\;;\quad
\frac{d}{ds'}r_v^{(0)}(\ve^2z;s,\cdot)\bigg|_{s'=\pm1}=0\,;
\ee
and we also used the fact that 
\be
\frac{d^2}{d{s'}^2}r_v^{(0)}(\ve^2z;s,s')=-\ve^2zr_v^{(0)}(\ve^2z;s,s') -\de(s-s')\,.
\ee
We remark that $r_v^{(0)}(z;s,s')=r_v^{(0)}(z;s',s)$, see equation \eqref{rv03}. We can rewrite the equation \eqref{225} as 
\be
\begin{aligned}
r_v(\ve^2 z;s,-1)
=
&
r_v^{(0)}(\ve^2 z;s,-1)-
\bigg[r_v^{(0)}(\ve^2z)(-\ga^2/4)\bigg(r_v(\ve^2 z;\cdot,-1)+\frac{y^*y^*(-1)}{\ve^2z}\bigg)\bigg](s)\\
&-\big(\ve^2z[r_v^{(0)}(\ve^2z)y^*](s)+y^*(s)\big)     \frac{y^*(-1)}{\ve^2z}
\end{aligned}
\ee
or, equivalently.
\be
\begin{aligned}
&r_v(\ve^2 z;s,-1)+y^*(s)\frac{y^*(-1)}{\ve^2z}\\
=\,
&
r_v^{(0)}(\ve^2 z;s,-1)-
\bigg[r_v^{(0)}(\ve^2z)(-\ga^2/4)\bigg(r_v(\ve^2 z;\cdot,-1)+\frac{y^*y^*(-1)}{\ve^2z}\bigg)\bigg](s)\\
&-\big[r_v^{(0)}(\ve^2z)y^*y^*(-1)\big](s)\,.
\end{aligned}
\ee
Now we can give an estimate of the derivative of the latter equation 
\be
\begin{aligned}
&\bigg\|\frac{d}{ds}\big[r_v(\ve^2 z;\cdot,-1)\big]+\frac{{y^*}'y^*(-1)}{\ve^2z}\bigg\|_{L^2((-1,1))}
\\
\leqslant\,
&
\bigg\|\frac{d}{ds}r_v^{(0)}(\ve^2 z;\cdot,-1)\bigg\|_{L^2((-1,1))}
+
\bigg\|\frac{d}{ds}\bigg[r_v^{(0)}(\ve^2z)(-\ga^2/4)\bigg(r_v(\ve^2 z;\cdot,-1)+\frac{y^*y^*(-1)}{\ve^2z}\bigg)\bigg]\bigg\|_{L^2((-1,1))}
\\
&
+\bigg\|\frac{d}{ds}\bigg[r_v^{(0)}(\ve^2z)y^*y^*(-1)\bigg]\bigg\|_{L^2((-1,1))}\leqslant c\,,
\end{aligned}
\ee
where we used the triangle inequality, estimates \eqref{jazz} and \eqref{ddsrv0}, and the fact that $\ga$ is bounded. Moreover we also used the result stated in Remark \ref{l2normrv} and the fact that $\|y^*\|_{L^2((-1,1))}=1$.  The proof of the second estimate in equation \eqref{estddsrv2} is identical and we omit it.
\end{proof}

\begin{proposition}
\label{prop-estpsiv}
Let $\hatpsiv$ be as in equation \eqref{psiv}. Then there exist $\ve_0>0$ such that for all $0<\ve<\ve_0$ the following estimates hold true:
\begin{itemize}
\item[\emph{Case 1}]
\beq
\label{estpsiv1}
\|\hatpsiv\|_{L^2(V)}\leqslant c \ve (|\xiuu|+|\xidd|)\;,\quad
\bigg\|\pd{}{s}\hatpsiv\bigg \|_{L^2(V)}\leqslant c \ve (|\xiuu|+|\xidd|)\,;
\eeq
\item[\emph{Case 2}]
\beq
\label{estpsiv2}
\big\|\hatpsiv-\psistar\big\|_{L^2(V)}\leqslant c \ve (|\xiuu|+|\xidd|)\;,\quad
\bigg\|\pd{}{s}\bigg[\hatpsiv-\psistar\bigg]\bigg \|_{L^2(V)}\leqslant c \ve (|\xiuu|+|\xidd|)\,,
\eeq
with 
\be
\psistar(s,u)=-\frac{1}{\ve}\frac{y^*(s)}{z}[\xiuu y^*(-1)+\xidd y^*(1)]\chi_n(u)\,.
\ee
\end{itemize}
where $c$ is a constant which does not depend on $\ve$, $f_1$, $f_2$ and $n$.
\end{proposition}
\begin{proof}
The proposition is a direct consequence of the definition of $\hatpsiv$, see equation \eqref{psiv},   and of Remark \ref{l2normrv} and Proposition  \ref{estimatesrv}. 
\end{proof}

The following remark applies only to  \emph{Case 2}:
\begin{remark}
\label{exphatpsi*}
From Theorem \ref{asymbeh},  equation \eqref{xi1xi2asym2}, and since $y^*(-1)=\al_1$ and $y^*(1)=\al_2$ we have that there exists $\ve_0>0$ such that for all $0<\ve<\ve_0$
\be
\frac{1}\ve\big|\xiuu y^*(-1)+\xidd y^*(1)\big|\leqslant c (|p_1|+|p_2|)
\ee
where $c$ is a constant which does not depend on $\ve$, $p_1$ and $p_2$. Thus implying 
\beq
\label{stimapsi*}
\big\|\psistar\big\|_{L^2(V)}\leqslant  c (|p_1|+|p_2|)
\;;\quad
\bigg\|\pd{}{s}\psistar\bigg \|_{L^2(V)}\leqslant  c (|p_1|+|p_2|)\,.
\eeq
\end{remark}

The following two propositions make preparations for the proof of Theorem \ref{resconv}.

\begin{proposition}[Regularity properties of $\hatPsi$]
\label{regularity}
For any vector $\tilde{\Xi}_n\equiv(f_1\chi_n,f_2\chi_n,0)$ with $f_1,f_2\in L^2((0,\infty))$  let us take $\hatPsi$ as it was done  in Definition \ref{hatpsi12v}. Then $\hatPsi\in D(\tilde H_\deve)$.
\end{proposition}
\begin{proof}
From the definition of $r_v(z)$, see equation \eqref{rv3}, it follows that the functions $r_v(\ve^2 z; s,\pm1)$ are well defined and that, for any $\ve>0$, $r_v(\ve^2 z; \cdot,\pm1)\in H^2((-1,1))$. Then for any $0<\de\leqslant\ve\leqslant1$ and from the definition of $\hatpsiuu$, $\hatpsidd$  and $\hatpsiv$, see equations \eqref{psi1} - \eqref{psiv}, it follows that, $\|\hat\Psi\|_{\Hduering}<\infty$. 

It remains to prove that the functions  $\hatpsiuu$, $\hatpsidd$ and $\hatpsiv$ satisfy the boundary conditions given in the definition  of $D(\tilde H_\deve)$, see equation  \eqref{DtHdeve}. The boundary conditions in $u=0,1$ are trivially satisfied due to  the definition of the function $\chi_n(u)$, see equation \eqref{chin}. 

Since the costants $\quu$ and $\qdd$ are related to the constants $\xiuu$ and $\xidd$ through the relation \eqref{qxi}, one has that the functions $\hatpsiuu$, $\hatpsidd$ and $\hatpsiv$ satisfy the conditions:
\begin{align*}
\hatpsiv(-1,u)&=\quu\chi_n(u)=\hatpsiuu(0,u)\\
\hatpsiv(1,u)&=\qdd\chi_n(u)=\hatpsidd(0,u)\,.
\end{align*}
From the definition of $r_v(z)$, see equation \eqref{rv3}, one has 
\begin{align}
&\frac{d}{ds}r_v(z;s,1)=\frac{d}{ds}\frac{\zeta_v(z;s)\eta_v(z;1)}{\WW_v(z;1)}=
\frac{\zeta_v'(z;s)}{\zeta_v'(z;1)}
\label{pixies1}
\\
&\frac{d}{ds}r_v(z;s,-1)=\frac{d}{ds}\frac{\eta_v(z;s)\zeta_v(z;-1)}{\WW_v(z;-1)}
=-\frac{\eta_v'(z;s)}{\eta_v'(z;-1)}\,,
\label{pixies2}
\end{align}
where we used $\WW_v(z)= \WW_v(z;1)=\WW_v(z;-1)$ in equations \eqref{pixies1} and \eqref{pixies2} respectively. Since $\zeta_v'(z;-1)=0$ and $\eta_v'(z;1)=0$ the following relations hold true
\be
\frac{d}{ds}r_v(z;s,1)\bigg|_{s=1}=1\;;\quad 
\frac{d}{ds}r_v(z;s,-1)\bigg|_{s=-1}=-1\;;
\ee
\be
\frac{d}{ds}r_v(z;s,1)\bigg|_{s=-1}=\frac{d}{ds}r_v(z;s,-1)\bigg|_{s=1}=0\,.
\ee
Hence
\begin{align*}
\pd{}{s}\hatpsiuu(0,u)&=\big(r_0(z)f_1\big)'(0)\chi_n(u)+i\sqrt{z}\quu\chi_n(u)=
\xiuu\chi_n(u)=-\frac{1}{\ve}\pd{}{s}\hatpsiv(-1,u)
\\
\pd{}{s}\hatpsidd(0,u)&=\big(r_0(z)f_2\big)'(0)\chi_n(u)+i\sqrt{z}\qdd\chi_n(u)=
\xidd\chi_n(u)=\frac{1}{\ve}\pd{}{s}\hatpsiv(1,u)\,,
\end{align*}
and we have proven that $\hatPsi\in D(\tilde H_\deve)$.
\end{proof}

\begin{proposition}
\label{p:estL}
For any couple of constants $\xiuu$ and $\xidd$ let us take $\hatpsiv$ as we did in equation \eqref{psiv}, then 
\begin{itemize}
\item[\emph{Case 1.}]
\beq
\label{tieni1}
\bigg\|\bigg[\tilde{L}_\defve-\frac{n^2\pi^2}{(\de/\ve)^2}-\ve^2 z\bigg]\hatpsiv\bigg\|_{L^2(V)}\leqslant
c \de [|\xiuu|+|\xidd|]  \,;
\eeq
\item[\emph{Case 2.}]
\beq
\label{tieni2}
\bigg\|\bigg[\tilde{L}_\defve-\frac{n^2\pi^2}{(\de/\ve)^2}-\ve^2z \bigg]\hatpsiv\bigg\|_{L^2(V)}\leqslant
c \de [|\xiuu|+|\xidd|] + \tilde c \frac{\de}{\ve}\bigg[\big\|\psistar\big\|_{L^2(V)}+
\bigg\|\pd{}{s}\psistar\bigg \|_{L^2(V)}\bigg]\,;
\eeq
\end{itemize}
where $\tilde{L}_\defve$ was defined in \eqref{L} and $c$ and $\tilde c$ are two constants which do not depend on $\ve$, $f_1$, $f_2$ and $n$.
\end{proposition}
\begin{proof}
We first note that, due to the regularity properties of the function $\hatpsiv$, see Proposition \ref{regularity}, for any $0<\de\leqslant\ve\leqslant1$ the function  $\tilde{L}_\defve \hatpsiv$  belongs to $L^2(V)$. Then we note that by our assumption $\gamma\in C^\infty_0((-1,1))$ and from the definition of $g_\defve$ and $W_\defve$, see equations \eqref{gdefve} and \eqref{W} respectively,  one has the following estimates
\beq
\label{estWg}
\bigg\|\frac{1}{g_\defve}-1\bigg\|_{L^\infty(V)}\leqslant
c \frac{\de}{\ve}\;;\quad
\bigg\|W_\defve+\frac{\ga^2}{4}\bigg\|_{L^\infty(V)}\leqslant
c \frac{\de}{\ve}\;;\quad
\bigg\|\pd{}{s}\frac{1}{g_\defve}\bigg\|_{L^\infty(V)}\leqslant
c \frac{\de}{\ve}\,.
\eeq
From the triangle inequality it follows that 
\be
\begin{aligned}
&\bigg\|\bigg[\tilde{L}_\defve-\frac{n^2\pi^2}{(\de/\ve)^2}-\ve^2 z\bigg]\hatpsiv\bigg\|_{L^2(V)}\\
\leqslant&
\bigg\|
\bigg[-\frac{1}{g_\defve}\pd{^2}{s^2}-\frac{\ga^2}{4}-\ve^2
 z\bigg]\hatpsiv\bigg\|_{L^2(V)}\\
&+
\bigg\|W_\defve+\frac{\ga^2}{4}\bigg\|_{L^\infty(V)}\big\|\hatpsiv\big\|_{L^2(V)}+
\bigg\|\pd{}{s}\frac{1}{g_\defve}\bigg\|_{L^\infty(V)}
\bigg\|\pd{}{s}\hatpsiv\bigg\|_{L^2(V)}\,,
\end{aligned}
\ee
where we used the fact that 
\be
-\pd{^2}{u^2}\hatpsiv=n^2\pi^2\hatpsiv\,.
\ee
By noticing that 
\be
\bigg[-\frac{1}{g_\defve}\pd{^2}{s^2}-\frac{\ga^2}{4}-\ve^2 z\bigg]\hatpsiv=
\bigg[\frac{1}{g_\defve}-1\bigg]\bigg[\frac{\ga^2}{4}+\ve^2 z\bigg]\hatpsiv\,,
\ee
and using the estimates in equation \eqref{estWg} we get 
\be
\bigg\|\bigg[\tilde{L}_\defve-\frac{n^2\pi^2}{(\de/\ve)^2}-\ve^2 z\bigg]\hatpsiv\bigg\|_{L^2(V)}\\
\leqslant
c \frac{\de}{\ve}\bigg[
\big\|\hatpsiv\big\|_{L^2(V)}+
\bigg\|\pd{}{s}\hatpsiv\bigg\|_{L^2(V)}\bigg]\,.
\ee
In \emph{Case 1} the statement follows from Proposition \ref{prop-estpsiv}. In \emph{Case 2} one has to use the trivial inequalities 
\be
\begin{aligned}
&\big\|\hatpsiv\big\|_{L^2(V)}\leqslant \big\|\hatpsiv-\psistar\big\|_{L^2(V)}+\big\|\psistar\big\|_{L^2(V)}
\;;\\
\quad
&\bigg\|\pd{}{s}\hatpsiv\bigg \|_{L^2(V)}\leqslant 
\bigg\|\pd{}{s}\bigg[\hatpsiv-\psistar\bigg]\bigg \|_{L^2(V)}+\bigg\|\pd{}{s}\psistar\bigg\|_{L^2(V)}
\end{aligned}
\ee
and the statement \eqref{tieni2} follows from the estimates \eqref{estpsiv2}. 
\end{proof}

We are now ready to give  the proof of Theorem \ref{resconv}.
\begin{proof}[{\bf Proof of Theorem \ref{resconv}}]
The fact that $\hatPsi \in D(\tilde H_\deve )$ was proved in Proposition \ref{regularity}. From the definition of the Hamiltonian $\tilde H_\deve$ and of the vector $\hatPsi $ we have that 
\be
\bigg[\tilde H_\deve-\frac{n^2\pi^2}{\de^2}-z\bigg]\hatPsi=
\bigg(f_1\chi_n,f_2\chi_n,
\frac{1}{\ve^2}\bigg[\tilde{L}_\defve-\frac{n^2\pi^2}{(\de/\ve)^2}-\ve^2 z\bigg]\hatpsiv\bigg)\,,
\ee
where we used 
\be
\bigg[-\pd{^2}{s^2}-
 \frac{1}{\de^2}\pd{^2}{u^2}-\frac{n^2\pi^2}{\de^2}-z\bigg]\hatpsijj=f_j\chi_n\qquad j=1,2\,.
 \ee
Then 
\beq
\label{here}
\bigg\|\bigg[\tilde H_\deve-\frac{n^2\pi^2}{\de^2}-z\bigg]\hatPsi-\tilde{\Xi}_n\bigg\|_{\tilde \HH_\ve}=
\frac{1}{\ve^{3/2}}\bigg\|\bigg[\tilde{L}_\defve-\frac{n^2\pi^2}{(\de/\ve)^2}-\ve^2 z\bigg]\hatpsiv\bigg\|_{L^2(V)}\,,
\eeq
recall the definition of $\|\cdot\|_{\tilde \HH_\ve} $ in Eq.  \eqref{2.4a}.  We note moreover that from the definition of $p_1$ and $p_2$, see equation \eqref{hatxixi}, one has
\beq
\label{p1p2}
|p_j|\leqslant c\|f_j\|_{L^2((0,\infty))}\,;\quad j=1,2\,.
\eeq
Using the last estimate in equations \eqref{xi1xi2asym2} and \eqref{stimapsi*} it follows that
\be
[|\xiuu|+|\xidd|]\leqslant c\|\tilde{\Xi}_n\|_{\tilde\HH_\ve} \;;\quad
\bigg[\big\|\psistar\big\|_{L^2(V)}+
\bigg\|\pd{}{s}\psistar\bigg \|_{L^2(V)}\bigg]\leqslant  c\|\tilde{\Xi}_n\|_{\tilde\HH_\ve}\,,
\ee
where we used $ \|\tilde{\Xi}_n\|_{\tilde\HH_\ve} =\big[\|f_1\|^2_{L^2((0,\infty))}+\|f_2\|^2_{L^2((0,\infty))}\big]^{1/2}$ and where $c$ is a  constant which does not depend on $\ve$, $f_1$, $f_2$ and $n$. Then  Theorem \ref{resconv}  follows from the  last estimate, from equation \eqref{here} and from    Proposition \ref{p:estL}.
\end{proof}

\section{Limit operator on the graph}\label{sec:graph}

We denote by $\HH_\GG$ the complex Hilbert space 
\be
\HH_\GG:=L^2((0,\infty))\oplus L^2((0,\infty))
\ee
with standard scalar product and norm. In $\HH_\GG$ we define the following selfadjoint operators: 
\begin{definition}[Limit operators on the graph]
\label{opsonG}
\be
D(\hdec):=\{(x_1,x_2)\in\HH_\GG|x_1,x_2\in H^2((0,\infty));\,x_1(0)=x_2(0)=0\}
\ee
\be
\hdec(x_1,x_2)=(-x_1'',-x_2'')\,.
\ee
\[
D(\halpha):=\bigg\{(x_1,x_2)\in\HH_\GG\Big|\,x_1,x_2\in H^2((0,\infty));\;
\Lambda_0^\perp
\begin{pmatrix}
x_1(0)\\
x_2(0)
\end{pmatrix}=0
;\;
\Lambda_0
\begin{pmatrix}
x_1'(0)\\
x_2'(0)
\end{pmatrix}=0
\bigg\}
\]
\[
\halpha(x_1,x_2)=(-x_1'',-x_2'')\,,
\]
where $\Lambda_0$ is the projector defined in equation  \eqref{proj}.
\end{definition}
We note that the operators $\hdec$ and $\halpha$ belong to the family of operators $-\Delta_\GG^{\Pi,\Theta}$ described in the introduction. In particular the operator $\hdec$  coincides with the Dirichlet (or decoupling) Laplacian on the graph and the operator $\halpha$ coincides with the weighted Laplacian  (for $N=2$).

For any $z\in\CO\backslash\RE$ we denote by $\rdec(z)$ and $\ralpha(z)$ the resolvents of $\hdec$ and $\halpha$ respectively, $\rdec(z)=(\hdec-z)^{-1}$ and $\ralpha(z)=(\halpha-z)^{-1}$. We note that for any vector $(f_1,f_2)\in \HH_\GG$  
\be
\rdec(z)(f_1,f_2)=(r_0(z)f_1,r_0(z)f_2)\,,
\ee
where the operator $r_0(z)$ was defined in equation \eqref{r0z}.  Moreover
\be
\ralpha(z)(f_1,f_2)=(r_0(z)f_1+q_1e^{i\sqrt{z}\,\cdot},r_0(z)f_2+q_2e^{i\sqrt{z}\,\cdot})\,,
\ee
with 
\be
\begin{pmatrix}
q_1\\ \\
q_2
\end{pmatrix}
=\frac{i\Lambda_0}{\sqrt{z}}
\begin{pmatrix}
p_1\\ \\
p_2
\end{pmatrix}
\;;\quad 
\puu=\big(r_0(z)f_1\big)'(0)\,;\;\pdd:=\big(r_0(z)f_2\big)'(0)\,;
\ee
note that   $\puu$ and $\pdd$ are defined accordingly to the definition used in Eq.  \eqref{hatxixi}.\\

Let us define the operator $\tilde{\PP}_n:\tilde \HH_\ve\to\HH_\GG$ 
\be
\tilde{\PP}_n\tilde\Psi:=\big((\chi_n,\tilde\psi_1)_{L^2((0,1))},(\chi_n,\tilde\psi_2)_{L^2((0,1))}\big)\in \HH_\GG\,,
\ee
for any vector $\tilde \Psi\equiv (\tilde\psi_1,\tilde\psi_2,\tilde\psi_v)\in\tilde \HH_\ve$. We denote by  $\tilde{\PP}_n^*:\HH_\GG\to \tilde \HH_\ve$ the adjoint of $\tilde{\PP}_n$. For any vector $(g_1,g_2)\in\HH_\GG$
 the action of $\tilde{\PP}_n^*$ is given by 
\be
\tilde{\PP}_n^*(g_1,g_2)=(g_1\chi_n,g_2\chi_n,0)\in\tilde\HH_\ve\,.
\ee

\begin{theorem}
\label{th3}
For any vector $\tilde{\Xi}_n\equiv(f_1\chi_n,f_2\chi_n,0)$, and  for all $z\in\CO\backslash\RE$  there exists $\ve_0>0$ such that for all $0<\ve<\ve_0$  the following estimates hold true: 
\begin{itemize}
\item[\emph{Case 1.}] For all $0<\de\leqslant\ve^{3/2}$
\be
\Big\|\Big[\tilde{\PP}_n\tilde R_{\deve}(z-n^2\pi^2/\de^2)-\rdec(z)\tilde{\PP}_n\Big]\tilde{\Xi}_n\Big\|_{\HH_\GG}\leqslant c
\bigg[\frac{\de}{\ve^{3/2}}  + \ve\bigg]  \|(f_1,f_2)\|_{\HH_\GG}\,;
\ee
\item[\emph{Case 2.}]For all $0<\de\leqslant\ve^{5/2}$

\be
\Big\|\Big[\tilde{\PP}_n\tilde R_{\deve}(z-n^2\pi^2/\de^2)-\ralpha(z)\tilde{\PP}_n\Big]\tilde{\Xi}_n\Big\|_{\HH_\GG}\leqslant
c\bigg[\frac{\de}{\ve^{5/2}} + \ve\bigg] \|(f_1,f_2)\|_{\HH_\GG}\,;
\ee
\end{itemize}
where $c$ is a constant which does not depend on $\ve$, $f_1$, $f_2$  and $n$, and  
\be
\tilde R_{\deve}(z-n^2\pi^2/\de^2):=\bigg(\tilde H_\deve-\frac{n^2\pi^2}{\de^2}-z\bigg)^{-1}\,.
\ee
\end{theorem}
\begin{proof}
We give the proof for the \emph{Case 2} only. The proof in the \emph{Case 1} is identical and we omit it. Since the operators $\tilde R(z-n^2\pi^2/\de^2)$ and $\ralpha(z)$ are bounded by $|\Im z|^{-1}$, it is enough to prove that for any vector $(g_1,g_2)\in \HH_\GG$ one has
\be
\Big|\Big((g_1,g_2),\Big[\tilde{\PP}_n\tilde R_{\deve}(z-n^2\pi^2/\de^2)-\ralpha(z)\tilde{\PP}_n\Big]\tilde{\Xi}_n\Big)_{\HH_\GG}\Big|\leqslant
c\bigg(\frac{\de}{\ve^{3/2}}+  \frac{\de}{\ve^{5/2}}\bigg)   \|(g_1,g_2)\|_{\HH_\GG}\|(f_1,f_2)\|_{\HH_\GG}  .
\ee
The following inequality holds true
\beq
\label{house}
\begin{aligned}
&\Big|\Big((g_1,g_2),\Big[\tilde{\PP}_n\tilde R_{\deve}(z-n^2\pi^2/\de^2)-\ralpha(z)\tilde{\PP}_n\Big]\tilde{\Xi}_n\Big)_{\HH_\GG}\Big|
\\
\leqslant&
\Big|\Big((g_1,g_2),\tilde{\PP}_n\Big[\tilde R_{\deve}(z-n^2\pi^2/\de^2)\tilde{\Xi}_n-\hatPsi\Big]\Big)_{\HH_\GG}\Big|+
\Big|\Big((g_1,g_2),\Big[\tilde{\PP}_n\hatPsi-\ralpha(z)\tilde{\PP}_n\tilde{\Xi}_n\Big]\Big)_{\HH_\GG}\Big|\,,
\end{aligned}
\eeq
where the vector $\hatPsi $ was given in Def. \ref{hatpsi12v}. From Rem. \ref{remark} and Th. \ref{resconv}, and since $\|\tilde{\Xi}_n\|_{\tilde\HH_\ve} = \|(f_1,f_2)\|_{\HH_\GG} $, it follows that 
\be
\begin{aligned}
&\Big|\Big((g_1,g_2),\tilde{\PP}_n\Big[\tilde R_{\deve}(z-n^2\pi^2/\de^2)\tilde{\Xi}_n-\hatPsi\Big]\Big)_{\HH_\GG}\Big|
\\
=&
\Big|\Big(\tilde{\PP}_n^*(g_1,g_2),\Big[\tilde R_{\deve}(z-n^2\pi^2/\de^2)\tilde{\Xi}_n-\hatPsi\Big]\Big)_{\tilde \HH_\ve}\Big|\leqslant 
c\frac{\de}{\ve^{5/2}}\|(g_1,g_2)\|_{\HH_\GG}\|(f_1,f_2)\|_{\HH_\GG} .
\end{aligned}
\ee
The second term on the right hand side of equation \eqref{house} can be  explicitly written as
\[\begin{aligned}
&\Big|\Big((g_1,g_2),\Big[\tilde{\PP}_n\hatPsi-\ralpha(z)\tilde{\PP}_n\tilde{\Xi}_n\Big]\Big)_{\HH_\GG}\Big| \\ 
=&
\Big|\Big((g_1,g_2),\big([\quu-q_1]e^{i\sqrt{z}\,\cdot},[\qdd-q_2]e^{i\sqrt{z}\,\cdot}\big)\Big)_{\HH_\GG}\Big|
\leqslant c \ve \|(g_1,g_2)\|_{\HH_\GG}\|(f_1,f_2)\|_{\HH_\GG} ,
\end{aligned}\]
where we used the result of Theorem \ref{asymbeh}, equation \eqref{q1q2asym2}, and the estimate \eqref{p1p2}.
\end{proof}

\section{Conclusions}
\label{conc}

We used the unitary map $U_\deve$, see equation \eqref{Uv} and  studied
the problem in the Hilbert space $\tilde \HH_\ve$. In this section we
discuss our results in the more natural Hilbert space $\HH_\deve$.

In our model the  width of the waveguide $\de$ must be intended as a
function of $\ve$ such that $0<\de(\ve)<\ve^a$, for some positive, large 
enough constant $a$. 
In what follows we shall not mark explicitly the fact that $\de$
depends on $\ve$. 

Let use define the functions  $\chide$  
\be
\chide(u):=\de^{-1/2}\chi_n(u)=\sqrt{\frac{2}{\de}}\sin{n\pi u}\qquad n=\NN\,,
\ee
were the functions  $\chi_n$ were defined in equation \eqref{chin}.

The main idea of our approach  is to find an approximate solution for the resolvent equation, i.e., for any vector in $\HH_\deve$ of the form $\Xi_{\de,n}=(f_1\chi_{\de,n}, f_2\chi_{\de,n},0)$, where $f_1,f_2\in L^2((0,\infty))$ we look for  an approximate solution of the equation 
\[
\bigg[H_\deve-\frac{n^2\pi^2}{\de^2}-z\bigg]\Psi=\Xi_{\de,n}\qquad z\in\CO\backslash\RE\,.
\]
In section \ref{sec2} we showed an explicit approximate solution which is in the domain of the Hamiltonian $\tilde H_{\deve}$. The following proposition is a consequence of the fact that $U_\deve$ is unitary and maps $D(H_\deve)$ to $D(\tilde H_\deve)$, the proof follows  directly from Theorem \ref{resconv}.
\begin{proposition}
For any vector $\Xi_{\de,n}\equiv(f_1\chi_{\de,n},f_2\chi_{\de,n},0)$
 with $f_1,f_2\in L^2((0,\infty))$  let us take
 $\Psi_\ve=U_\deve^{-1}\hatPsi$, where the vector $\hatPsi$ was given in
 Definition \ref{hatpsi12v}. Moreover in
 Case 1 assume $\de(\ve)<\ve^{3/2}$ and in Case 2 
 assume $\de(\ve)<\ve^{5/2}$. Then $\Psi_\ve\in D(H_\deve)$, and for all
 $z\in\CO\backslash\RE$ 
\[
\bigg[H_\deve-\frac{n^2\pi^2}{\de^2}-z\bigg]\Psi_\ve=\Xi_{\de,n}+\Phi_\ve
\]
with 
\be
\|\Phi_\ve\|_{\HH_\deve}\leqslant c_\ve \|\Xi_{\de,n}\|_{\HH_\deve} 
\,,
\ee
where $c_\ve$ does not depend on  $f_1$, $f_2$ and $n$, and $c_\ve\to 0$ as $\ve\to 0$.
\end{proposition}
To prove the convergence to an operator on the graph (in norm resolvent
sense) we note that  in the edges the approximate solution $\Psi_\ve = U_\deve^{-1}\hatPsi$ remains  factorized in the coordinates $s$ and $u$: 
\[
\begin{aligned}
&\psi_{1,\ve}(s,u)=\big[\big(r_0(z)f_1\big)(s)+\quu e^{i\sqrt{z}s}\big]\chi_{\de,n}(u)
\equiv x_{1,\ve}(s)\chi_{\de,n}(u)\\
&\psi_{2,\ve}(s,u)=\big[\big(r_0(z)f_2\big)(s)+\qdd e^{i\sqrt{z}s}\big]\chi_{\de,n}(u)
\equiv x_{2,\ve}(s)\chi_{\de,n}(u)\,,
\end{aligned}
\]
see also Def. \ref{hatpsi12v}. We interpret the $s$ dependent parts in $\psi_{1,\ve}$ and $\psi_{2,\ve}$, i.e., the functions $x_{1,\ve}$ and $x_{2,\ve}$,  as functions on the edges of the graph and consider the  vector $(x_{1,\ve},x_{2,\ve})\in\HH_\GG$. By definition for all $\ve>0$, we have $x_{1,\ve},\,x_{2,\ve}\in H^2((0,\infty))$ and these functions  satisfy the equation 
\be
-x_{j,\ve}''-zx_{j,\ve}=f_j\qquad j=1,2\,.
\ee
As we are  able to
compute the  limit of $x_{\ve,1}(0)\equiv \quu$,
$x_{\ve,2}(0)\equiv\qdd$, $x_{\ve,1}'(0)\equiv\xiuu$ and
$x_{\ve,2}'(0)\equiv\xidd$ as $\ve\to0$,  we can  prove that, see
Theorem \ref{asymbeh},\\ 
in \emph{Case 1}
\be
\lim_{\ve\to0}x_{\ve,1}(0) =0\,,\;
\lim_{\ve\to0}x_{\ve,2}(0)=0 \,,\;
\lim_{\ve\to0}x_{\ve,1}'(0)=p_1\,,\; 
\lim_{\ve\to0}x_{\ve,2}'(0)=p_2\,,
\ee
implying  that the limit operator on the graph is the Laplacian with Dirichlet conditions in the vertex;\\
while in \emph{Case 2}
\[
\lim_{\ve\to0}\big[\al_2 x_{\ve,1}(0)-\al_1x_{\ve,2}(0)\big]=0\,,\;
\lim_{\ve\to0}\big[\al_1x_{\ve,1}'(0)+\al_2x_{\ve,2}'(0)\big]=0 
\]
where $\al_1$ and $\al_2$ are two real constants defined by the zero energy eigenvector $y^*$ of the Hamiltonian $h_v$, see equations \eqref{eqv*} and \eqref{al1al2}. In this case the limit operator on the graph is the Laplacian with a weighted Kirchhoff condition in the vertex. In  \emph{Case 2} one of the constants $\al_1$ or $\al_2$ may be equal to zero but they cannot be both equal to zero; in this case  the limit operator on the graph is defined by a Dirichlet condition on one of the edges and a Neumann condition on the other.
 
We remark that the analysis performed here can be used also to prove the results of \cite{cacciapuoti-exner:07} and \cite{cacciapuoti-finco:aa10} which cover some generalizations of the model presented in \cite{albeverio-cacciapuoti-finco:07}. 

We also note that our method applies to settings with one or both
edges having finite length. The form of the solution in the vertex
region (i.e. the function $\hatpsiv$ in Definition \ref{hatpsi12v}) does
not change, while the solution in the edge regions must be adapted to fulfill the boundary condition in each  endpoint.
\\

{\bf Acknowledgments.} The author is grateful to Gianfausto Dell'Antonio
and Emanuele Costa for many enlightening exchanges of views. The author
is also indebted to Sergio Albeverio for pointing out at first the topic
of the paper to his attention, for useful comments and discussions,  and
for the kind support during the writing of this work. The warm
hospitality and support of the Mathematical Institute 
of Tohoku
University are also  gratefully acknowledged. Most of the  work  was done while the author was employed at the
Hausdorff Institute for Mathematics which is acknowledged for the
support. The work was also partially financed by the JSPS postdoctoral
fellowship program and by the FIR 2013 project ``Condensed Matter in Mathematical Physics'', Ministry of University and
Research of Italian Republic  (code RBFR13WAET).


\begin{thebibliography}{10}

\bibitem{albeverio-cacciapuoti-finco:07}
Albeverio, S., Cacciapuoti, C., and Finco, D., \emph{Coupling in the singular
  limit of thin quantum waveguides}, J. Math. Phys. \textbf{48} (2007), 032103.

\bibitem{albeverio-kusuoka:pp10}
Albeverio, S. and Kusuoka, S., \emph{Diffusion processes in thin tubes and
  their limits on graphs}, Ann. Probab. \textbf{40} (2012), no. 4, 2131--2167.

\bibitem{berkolaiko-carlson-fulling-kuchment:06}
Berkolaiko, G., Carlson, R., Fulling, S., and Kuchment, P., \emph{Quantum
  graphs and their applications}, Contemporary Math., vol. 415, Am. Math. Soc.,
  Providence, RI, 2006.

\bibitem{berkolaiko-kuchment:13}
Berkolaiko, G. and Kuchment, P., \emph{Introduction to quantum graphs},
  Mathematical Surveys and Monographs, vol. 186, Am. Math. Soc., Providence,
  RI, 2013.

\bibitem{bonciocat:08}
Bonciocat, A.~I., \emph{Curvature bounds and heat kernels: discrete versus
  continuous spaces}, Ph.D. Thesis, Universit{\"a}t Bonn,
  \url{http://hss.ulb.uni-bonn.de:90/2008/1497/1497.htm}, 2008.

\bibitem{bouchitte-mascarenhas-trabucho:esaim07}
Bouchitt\'{e}, G., Mascarenhas, M.~L., and Trabucho, L., \emph{On the curvature
  and torsion effects in one dimensional waveguides}, ESAIM: Control Optim.
  Calc. Var. \textbf{13} (2007), no.~4, 793--808.

\bibitem{cacciapuoti-exner:07}
Cacciapuoti, C. and Exner, P., \emph{Nontrivial edge coupling from a
  {D}irichlet network squeezing: the case of a bent waveguide}, J. Phys. A:
  Math. Theor. \textbf{40} (2007), no.~26, F511--F523.

\bibitem{cacciapuoti-finco:aa10}
Cacciapuoti, C. and Finco, D., \emph{Graph-like models for thin waveguides with
  {R}obin boundary conditions}, Asymptot. Anal. \textbf{70} (2010), 199--230.

\bibitem{cheon-exner-turek:ap10}
Cheon, T., Exner, P., and Turek, O., \emph{Approximation of a general singular
  vertex coupling in quantum graphs}, Ann. Physics \textbf{325} (2010), no.~3,
  548--578.

\bibitem{deverdiere:86}
Colin~de Verdiere, Y., \emph{Sur la multiplicit{\'e} de la premiere valeur
  propre non nulle du laplacien}, Commentarii Mathematici Helvetici \textbf{61}
  (1986), no.~1, 254--270.

\bibitem{oliveira:rxv10}
{De Oliveira}, C.~R., \emph{Quantum singular operator limits of thin
  {D}irichlet tubes via {$\Gamma$}-convergence}, arXiv:1010.2101v1 [math-ph]
  (2010), 28pp.

\bibitem{dellantonio-costa:10}
Dell'Antonio, G.~F. and Costa, E., \emph{Effective {S}chr\"{o}dinger dynamics
  on $\varepsilon$-thin {D}irichlet waveguides via quantum graphs: {I}.
  {S}tar-shaped graphs}, J. Phys. A: Math. Theo. \textbf{43} (2010), no.~47,
  474014, 23pp.

\bibitem{dellantonio-michelangeli:15}
Dell'Antonio, G. and Michelangeli, A., \emph{Dynamics on a graph as the limit
  of the dynamics on a ``fat graph''}, Mathematical Technology of Networks,
  Springer, 2015, pp.~49--64.

\bibitem{dellantonio-tenuta:jmp06}
Dell'Antonio, G. and Tenuta, L., \emph{Quantum graphs as holonomic
  constraints}, J. Math. Phys. \textbf{47} (2006), 072102.

\bibitem{duclos-exner:95}
Duclos, P. and Exner, P., \emph{Curvature-induced bound states in quantum
  waveguides in two and three dimensions}, Rev. in Math. Phys. \textbf{7}
  (1995), no.~1, 73--102.

\bibitem{exner-keating-kuchment-sunada-teplyaev:08}
Exner, P., Keating, J.~P., Kuchment, P., Sunada, T., and Teplyaev, A.,
  \emph{Analysis on graphs and its applications}, Proceedings of Symposia in
  Pure Mathematics, vol.~77, Am. Math. Soc., Providence, RI, 2008.

\bibitem{exner-kovarik:15}
Exner, P. and Kova\v{r}\'{i}k, H., \emph{Quantum waveguides}, Theoretical and
  Mathematical Physics, Springer, 2015.

\bibitem{exner:11}
Exner, P., \emph{Vertex couplings in quantum graphs: approximations by scaled
  {S}chr\"odinger operators}, Proceedings of the ICM Satellite Conference
  Mathematics in Science and Technology (New Delhi 2010), 2011, pp.~71--92.

\bibitem{exner-post:05}
Exner, P. and Post, O., \emph{Convergence of spectra of graph-like thin
  manifolds}, J. Geom. Phys. \textbf{54} (2005), 77--115.

\bibitem{exner-post:07-1}
Exner, P. and Post, O., \emph{Convergence of resonances on thin branched
  quantum waveguides}, J. Math. Phys. \textbf{48} (2007), 092104, 43pp.

\bibitem{exner-post:09}
Exner, P. and Post, O., \emph{Approximation of quantum graph vertex couplings
  by scaled {S}chr{\"o}dinger operators on thin branched manifolds}, J. Phys.
  A: Math. Theo. \textbf{42} (2009), 415305, 22pp.

\bibitem{ESjmp}
Exner, P. and \v{S}eba, P., \emph{Bound states in curved quantum waveguides},
  J. Math. Phys. \textbf{30} (1989), no.~11, 2574--2580.

\bibitem{freidlin-wentzel:93}
Freidlin, M.~I. and Wentzel, A.~D., \emph{Diffusion processes on graphs and
  averaging principle}, Ann. Probab. \textbf{21} (1993), no.~4, 2215--2245.

\bibitem{golovaty-hryniv:09}
Golovaty, {\relax Yu D.}. and Hryniv, R.~O., \emph{On norm resolvent
  convergence of {S}chr{\"o}dinger operators with $\delta'$-like potentials},
  J. Phys. A: Math. Theor. \textbf{43} (2010), no.~15, 155204.

\bibitem{grieser:07}
Grieser, D., \emph{Spectra of graph neighborhoods and scattering}, Proc. London
  Math. Soc. \textbf{97} (2008), no.~3, 718--752.

\bibitem{harmer-pavlov-yafyasov:07}
Harmer, M., Pavlov, B., and Yafyasov, A., \emph{Boundary conditions at the
  junction}, J. Comput. Electron. \textbf{6} (2007), 153--157.

\bibitem{Hel67}
Hellwig, G., \emph{Differential operators of mathematical physics},
  Addison-Wesley Publ. Comp., Reading (MA) Palo Alto London Don Mills (ON),
  1967.

\bibitem{kato:80}
Kato, T., \emph{Perturbation theory for linear operators}, Springer-Verlag,
  Berlin Heidelberg New York, 1980.

\bibitem{kostrykin-schrader:99}
Kostrykin, V. and Schrader, R., \emph{{K}irchhoff's rule for quantum wires}, J.
  Phys. A: Math. Gen. \textbf{32} (1999), no.~4, 595--630.

\bibitem{kosugi:00}
Kosugi, S., \emph{A semilinear elliptic equation in a thin network-shaped
  domain}, J. Math. Soc. Japan \textbf{52} (2000), no.~3, 673--697.

\bibitem{kuchment:04}
Kuchment, P., \emph{Quantum graphs. {I}. {S}ome basic structures}, Waves Random
  Media \textbf{14} (2004), no.~1, S107--S128.

\bibitem{kuchment-zeng:01}
Kuchment, P. and Zeng, H., \emph{Convergence of spectra of mesoscopic systems
  collapsing onto a graph}, J. Math. Anal. Appl. \textbf{258} (2001), no.~2,
  671--700.

\bibitem{kuroda:11}
Kuroda, H., \emph{The {M}osco convergence of {D}irichlet forms approximating
  the {L}aplace operators with the delta potential on thin domains},
  arXiv:1106.5476 [math.AP] (2011), 16pp.

\bibitem{kuwae-shioya:03}
Kuwae, K. and Shioya, T., \emph{Convergence of spectral structures: a
  functional analytic theory and its applications to spectral geometry}, Comm.
  Anal. Geom. \textbf{11} (2003), no.~4, 599--673.

\bibitem{manko:12}
Man'ko, S.~S., \emph{Schr\"odinger operators on star graphs with singularly
  scaled potentials supported near the vertices}, J. Math. Phys. \textbf{53}
  (2012), no.~12, 123521, 13pp.

\bibitem{molchanov-vainberg:06-2}
Molchanov, S. and Vainberg, B., \emph{Laplace operator in networks of thin
  fibers: spectrum near the threshold}, Stochastic analysis in mathematical
  physics, Proceedings of a Satellite Conference of ICM 2006 (2006), 69--93,
  edited by G. Ben Arous, A.-B. Cruzeiro, Y. Le Jan, J.-C. Zambrini.

\bibitem{mugnolo:14}
Mugnolo, D., \emph{Semigroup methods for evolution equations on networks},
  Understanding Complex Systems, Springer, 2014.

\bibitem{post:12}
Post, O., \emph{Spectral analysis on graph-like spaces}, Lecture Notes in
  Mathematics, vol. 2039, Springer, 2012.

\bibitem{post:05}
Post, O., \emph{Branched quantum wave guides with {D}irichlet boundary
  conditions: the decoupling case}, J. Phys. A: Math. Gen. \textbf{38} (2005),
  no.~22, 4917--4931.

\bibitem{post:06}
Post, O., \emph{Spectral convergence of quasi-one-dimensional spaces}, Ann.
  Henri Poincar{\'e} \textbf{7} (2006), 933--973.

\bibitem{raugel:95}
Raugel, G., \emph{Dynamics of partial differential equations on thin domains},
  Dynamical systems, Lecture Notes in Math. \textbf{1609} (1995), 208--315.

\bibitem{rubinstein-schatzman:01}
Rubinstein, J. and Schatzman, M., \emph{Variational problems on multiply
  connected thin strips. {I}. {B}asic estimates and convergence of the
  {L}aplacian spectrum}, Arch. Ration. Mech. Anal. \textbf{160} (2001), no.~4,
  271--308.

\bibitem{saito:00}
Sait\={o}, Y., \emph{The limiting equation for {N}eumann {L}aplacians on
  shrinking domains}, Electron. J. Diff. Equations \textbf{2000} (2000),
  no.~31, 1--25.

\bibitem{saito:01}
Sait\={o}, Y., \emph{Convergence of the {N}eumann {L}aplacian on shrinking
  domains}, Analysis (Munich) \textbf{21} (2001), no.~2, 171--204.

\bibitem{Tit62}
Titchmarsh, E.~C., \emph{Eigenfunction expansions: Part {I}}, Oxford University
  Press, London, 1962.

\bibitem{ueker-ea:15}
Uecker, H., Grieser, D., Sobirov, Z., Babajanov, D., and Matrasulov, D.,
  \emph{Soliton transport in tubular networks: transmission at vertices in the
  shrinking limit}, Phys. Rev. E \textbf{91} (2015), no.~2, 023209, 8pp.

\end{thebibliography}
%

\end{document}